\documentclass[12pt,preprint]{aastex}
\newcommand{\etal}{\mbox{\rm et al.~}}
\newcommand{\ms}{\mbox{m s$^{-1}~$}}
\newcommand{\kms}{\mbox{km s$^{-1}~$}}
\newcommand{\kmse}{\mbox{km s$^{-1}$}}

\newcommand{\mse}{\mbox{m s$^{-1}$}}

\newcommand{\msun}{M$_{\odot}~$}
\newcommand{\msune}{M$_{\odot}$}

\newcommand{\rsune}{R$_{\odot}$}
\newcommand{\lsun}{L$_{\odot}~$}
\newcommand{\lsune}{L$_{\odot}$}
\newcommand{\mjup}{M$_{\rm Jup}~$}
\newcommand{\mjupe}{M$_{\rm Jup}$}

\newcommand{\mnep}{M$_{\rm Nep}~$}

\newcommand{\mearth}{M$_{\rm Earth}~$}
\newcommand{\mearthe}{M$_{\rm Earth}$}

\newcommand{\msini}{$M \sin i~$}
\newcommand{\msinie}{$M \sin i$}
\newcommand{\vsini}{$v \sin i~$}
\newcommand{\vsinie}{$v \sin i$}

\newcommand{\chisq}{$\chi_{\nu}^2~$}
\newcommand{\chisqe}{$\chi_{\nu}^2$}
\newcommand{\cs}{$\sqrt{\chi^2_{\nu}}~$}
\newcommand{\cse}{$\sqrt{\chi^2_{\nu}}$}

\newcommand{\rphk}{\ensuremath{R'_{\mbox{\scriptsize HK}}}}

\newcommand{\caii}{\ion{Ca}{2} H \& K~}

\received{}
\accepted{}

\slugcomment{ }

\shortauthors{Fischer {\it et~al.\/}}
\shorttitle{Five Extrasolar Planets}
\begin{document}

\title{Five Planets Orbiting 55 Cancri$^{1}$}
\author{Debra A. Fischer\altaffilmark{2},
        Geoffrey W. Marcy\altaffilmark{3}, 
        R. Paul Butler\altaffilmark{4},
        Steven S. Vogt\altaffilmark{5},
        Greg Laughlin\altaffilmark{5},
        Gregory W. Henry\altaffilmark{6},
	David Abouav\altaffilmark{2},
        Kathryn M. G. Peek\altaffilmark{3},
        Jason T. Wright\altaffilmark{3},
        John A. Johnson\altaffilmark{3},
        Chris McCarthy\altaffilmark{2},
        Howard Isaacson\altaffilmark{2}
}

\email{fischer@stars.sfsu.edu}

\altaffiltext{1}{Based on observations obtained at 
    the W.M. Keck Observatory, which is operated jointly
    by the University of California and the California
    Institute of Technology.  Keck time has been granted by
    both NASA and the University of California.}

\altaffiltext{2}{Department of Physics and Astronomy, 
San Francisco State University, San Francisco, CA, USA 94132}

\altaffiltext{3}{Department of Astronomy, University of California,
Berkeley, CA USA  94720-3411}

\altaffiltext{4}{Department of Terrestrial Magnetism, Carnegie Institution of
Washington, 5241 Broad Branch Rd NW,  Washington DC, USA 20015-1305}

\altaffiltext{5}{UCO/Lick Observatory, 
University of California at Santa Cruz, Santa Cruz, CA, USA 95064}

\altaffiltext{6}{Center of Excellence in Information Systems,
Tennessee State University, 3500 John A. Merritt Blvd, Box 9501,
Nashville, TN 37209}

\begin{abstract}
We report 18 years of Doppler shift measurements of a nearby star, 55
Cancri, that exhibit strong evidence for five orbiting planets.  The
four previously reported planets are strongly confirmed here.  A
fifth planet is presented, with an apparent orbital period of
260 days, placing it 0.78 AU from the star in the large empty zone 
between two other planets.  The velocity wobble amplitude of 4.9 \ms 
implies a minimum planet mass \msini = 45.7 \mearthe. The orbital 
eccentricity is consistent with a circular orbit, but modest 
eccentricity solutions give similar \chisq fits. 
All five planets reside in 
low eccentricity orbits, four having eccentricities under 0.1. The 
outermost planet orbits 5.8 AU from the star and has a minimum mass, 
\msini = 3.8 \mjupe, making it more massive than the
inner four planets combined.  Its orbital distance is the largest for
an exoplanet with a well defined orbit.  The innermost planet has a
semi-major axis of only 0.038 AU and has a minimum mass, \msinie, of
only 10.8 \mearthe, one of the lowest mass exoplanets known. 
The five known planets within 6 AU define
a {\em minimum mass protoplanetary nebula} to compare with the
classical minimum mass solar nebula.  Numerical N-body simulations
show this system of five planets to be dynamically stable and show that the
planets with periods of 14.65 and 44.3 d are not in a mean-motion resonance.
Millimagnitude photometry during 11 years reveals no
brightness variations at any of the radial velocity periods,
providing support for their interpretation as planetary.

\end{abstract}



\section{Introduction}
\label{intro}

One of the first few detected exoplanets was 
a planetary companion to 55 Cnc \citep{Butler97}.
At the time, eight years of Doppler measurements 
from Lick Observatory revealed a 14.6-day wobble in 55 Cnc as it was
gravitationally perturbed by a Jupiter-mass planet.  Superimposed
on this 14.6-day Doppler periodicity was an additional trend 
showing clear curvature and indicating that 55 Cancri was host to a second
orbiting body, likely of planetary mass.

Additional Doppler measurements through 2002 uncovered the full
Doppler cycle with a period of $\sim$14 yr, caused by a planet with a
minimum mass \msini = 4 \mjup orbiting $\sim$5.5 AU from 55 Cnc 
\citep{Marcy_55cnc}.  This was the first giant planet found 
with an orbital radius similar to the giant planets in our solar system.  A third Doppler
periodicity of 44.3 days was also apparent in those data, indicating a
third Jupiter-mass planet in the system, this one orbiting 0.25 AU
from the star \citep{Marcy_55cnc}.  This was the second planetary
system found to have three planets, the first being that around
$\upsilon$ Andromedae \citep{Butler_upsand}.

By combining these velocities from Lick Observatory with over
100 precise Doppler measurements obtained in one year at the
Hobby-Eberly Telescope, along with measurements from the ``Elodie''
spectrometer at Haute Provence, \citet{McArthur04} identified a fourth
planet having a small minimum mass of \msini = 14 \mearth with an
orbital period of 2.8 d.  This planet was one of the first three
Neptune-mass planets discovered, along with the planets orbiting
GJ~436 \citep{Butler04, Maness07} and $\mu$ Arae \citep{Santos04b}.
The detection of this Neptune-mass planet made 55 Cancri
the first system known to contain four planets.  \citet{McArthur04}
also used the Fine Guidance Sensor on the Hubble Space Telescope to
carry out astrometry of 55 Cnc, and estimated the inclination of the
orbital plane of the outer planet to be $i = 53 \pm 6.8$ deg.

This four-planet system left a large, dynamically empty gap between
0.25 and 6 AU. Numerical simulations suggested that hypothetical
planets in this gap would be dynamically stable, including the
interesting possibility of terrestrial mass planets in the habitable
zone between 0.5 and $\sim$2 AU \citep{Marcy_55cnc, Raymond06a}.

In 2004, we noticed modest peaks in the periodogram at 260 and 470 d, 
indicating possible planets at those periods that motivated our 
continued intense Doppler observations. 
\citet{Wisdom05a} carried out an independent analysis of the combined
published Doppler measurements and identified a 260 d periodicity, implying a new planet
with a minimum mass of 1.8 \mnep = 31 \mearthe.  It is not uncommon
for modest peaks in the periodogram to fluctuate in their confidence level
with the addition of new data, especially for cases where the radial velocity amplitudes 
are comparable to the precision of Doppler measurements, so we intensified 
our observations of this star.   
Here, we add 115 additional radial velocity measurements 
from Lick Observatory and 70 radial velocity measurements
from Keck Observatory to our previous data set and find that 
the false alarm probability for a 260-d signal has strengthened 
and is present in data sets from both Observatories independently.

A stellar companion orbits 55 Cancri as well.  It is a 13th magnitude
M dwarf located roughly 1000 AU away, certainly bound to 55 Cancri A
as the radial velocities are nearly the same.  The occurrence,
dynamics, and final properties of planetary systems may well be
affected by such stellar companions, as indicated in observational
studies by \citet{Eggenberger04} and \citet{Raghavan06}.  Thus 55
Cancri offers a test of the effects of binary
companions on the architecture of complex planetary systems.

Spitzer Space Telescope results for 55 Cnc by \citet{Bryden06}
show that the observed 24 and 70 micron flux densities are comparable
to the predicted brightness of the stellar photosphere, indicating no
infrared excess above the errors.  The corresponding upper limit
to the fractional infrared luminosity is 8$\times10^{-6}$, or about 80
zodis.  A detectable scattered light disk was also ruled out by its
non-detection in HST NICMOS data by \citet{Schneider01}.

Here we provide Doppler measurements from both Lick and Keck
Observatories that significantly augment the 2002 set from Lick
Observatory alone.  These measurements span a longer time
baseline and contain higher Doppler precision with the addition of new
Keck velocities, offering a chance to reassess all of the planets 
around 55 Cancri.

\section{Properties of 55 Cancri}

55 Cnc (=HD 75732=$\rho^1$ Cnc A = HR3522=HIP 43587) has an apparent
brightness $V = 5.96$ and Hipparcos parallax of $79.8 \pm 0.84$ mas
\citep{ESA97}, implying a distance of $12.5 \pm 0.13$ parsecs and
absolute visual magnitude $M_V = 5.47$.  Using spectra from the
California \& Carnegie Planet Search, \citet{Valenti05} derived
$T_{eff} = 5234 \pm 30$K, $\log g = 4.45 \pm 0.08$, $v \sin i = 2.4
\pm 0.5$\kms, and [Fe/H]$ = +0.31 \pm 0.04$.  Indeed, 55 Cnc is so
metal-rich as to be in the fifth metallicity percentile of stars
within 25 pc \citep{Valenti05}.  Using a bolometric correction that
accounts for the high metallicity \citep{VandenBerg03}, we calculate a
stellar luminosity of 0.6 \lsune.  The effective temperature,
spectroscopic surface gravity and intrinsic luminosity are all
consistent with a spectral classification of this star as K0/G8V.  The
star is chromospherically inactive, with a Mt.~Wilson $S$-value of
0.22 (averaged during the past seven years of our measurements)
implying $\log{\rphk} = -4.84$, indicating a modest age of 2-8 Gyr
where 2 Gyr is a strong lower limit on age.  The rotation period,
calibrated to this chromospheric activity, is estimated to be 39 days
\citep{Noyes84, Wright04a}.  However, for metal-rich stars,
chromospheric emission at the \caii lines remain poorly calibrated as
a diagnostic of rotation and age.  A more complete discussion of the
chromospheric activity and implied stellar properties are given by
\citet{Marcy_55cnc}.

\begin{figure}[t!]
\plotone{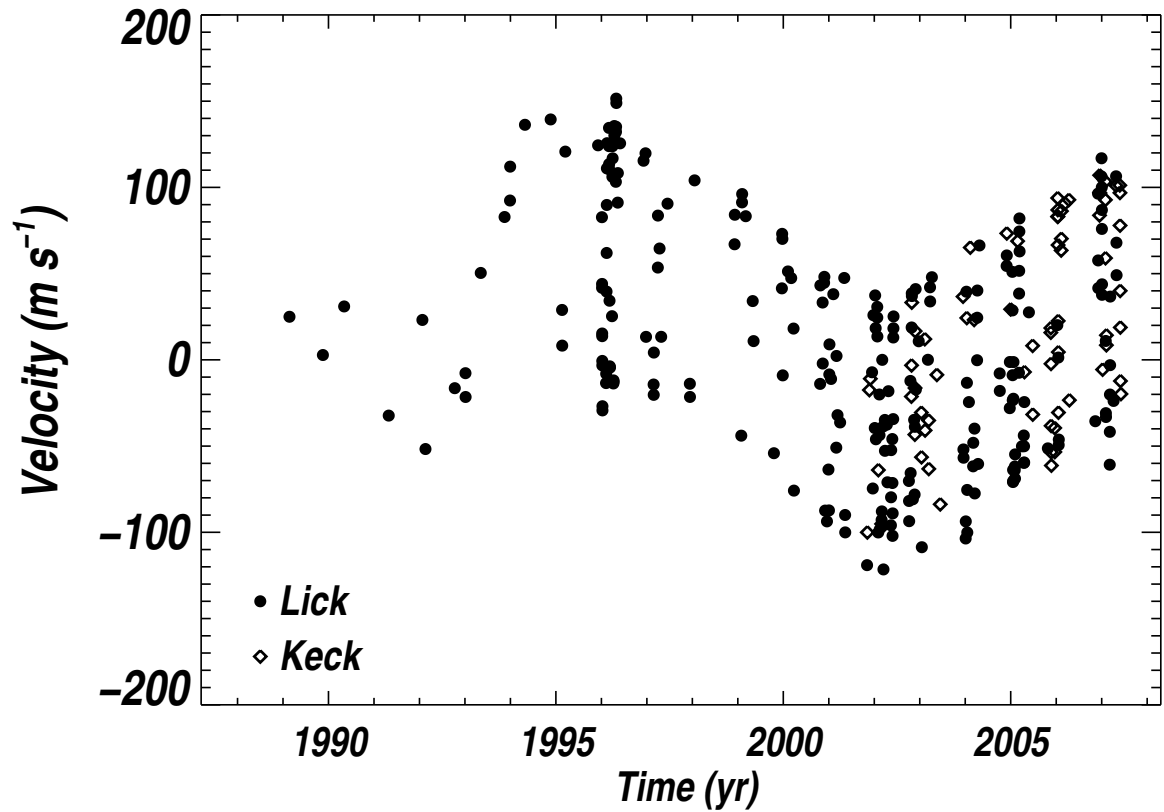}
\caption{Measured velocities for 55 Cancri from Lick and Keck obtained from
  1989.1 to 2007.4.  Data from Lick (filled dots) had errors of
  $\sim$10 \ms prior to 1994 and 3--5 \ms thereafter.  Data from Keck
  (open diamonds) had errors of $\sim$3 \ms prior to 2004 August, and
  1.0--1.5 \ms thereafter.  The 14-year period from the outer planet
  and the short timescale variations from the 14.6-day planet are
  apparent to the eye.}
\label{fig1}
\end{figure}

The mass of 55 Cnc is best determined by associating its measured
effective temperature, luminosity, and metallicity with models of
stellar interiors.  Using the well known ``Yale models,'' \citep{Yi04},
\citet{Valenti05} found a stellar mass of $0.92 \pm 0.05$ \msune.
\citet{Takeda07} have also derived modified stellar evolutionary
models using the Yale Stellar Evolution Code to match the observed
spectroscopic parameters from {\citet{Valenti05}. They derive a
stellar mass for 55 Cnc of $0.96 \pm 0.05$ \msun with the uncertainty
corresponding to the 99.7\% credibility intervals of the Bayesian
posterior probability distributions.  Here we simply adopt the average
of these two estimates, giving $M = 0.94 \pm 0.05$ \msun for the mass
of 55 Cnc.  We note that the adopted uncertainty in stellar mass
implies fractional errors in the derived planetary masses of 8 percent
in addition to errors in the orbital parameters.

\section{Doppler--Shift Measurements}

We have obtained 636 observations of the G8 main-sequence star
55 Cnc over the past 18 years.  We generally obtain two or three 
consecutive observations and bin them to increase the velocity 
precision and accuracy. Here, we present 
250 binned velocity measurements made at Lick Observatory from 1989-2007, and 70
binned velocity measurements made at the Keck Observatory from 2002--2007.  The Lick
spectra were obtained using both the 3-meter telescope and the
Coud\'{e} Auxiliary Telescope, which both feed the Hamilton optical
echelle spectrometer \citep{Vogt87}.  A detailed description of the
setup of the Hamilton spectrometer, its calibrating iodine absorption
cell, and the method of extracting Doppler measurements for 55 Cnc are
given in \citet{Butler96b} and in \citet{Marcy_55cnc}.  The Keck
spectra were obtained with the HIRES spectrometer \citep{Vogt94}, and
a description of that setup and Doppler measurements are given in
\citet{Butler06}.

At both telescopes, we place a cylindrical, pyrex cell filled with
molecular iodine gas in the light path of the telescope, just before
the spectrometer slit, to superimpose sharp absorption lines of known
wavelength on the stellar spectrum.  The iodine lines provide
calibration of both wavelength and the spectrometer PSF (Butler et al
1996).  While twenty percent of the starlight is absorbed by iodine, the 
cell's inclusion is worthwhile because the dense
iodine absorption lines provide a permanent record of the wavelength
scale and behavior of the spectrometer at the instant of each
observation, producing long-term Doppler precision free of systematic
errors to the level of 1 \mse.

The velocity measurements are listed in Table 1 (the full Table is 
available in the electronic version of this paper only) and shown in Figure
\ref{fig1} with different symbols for measurements made at Lick and
Keck.  We carried out a preliminary five-Keplerian fit to the combined velocities
from both telescopes, allowing one extra parameter to be the
difference in the velocity zero-point from the two spectrometers,
found to be $28.8 \pm 0.5$ \mse.  Once established, we applied the +28.8 \ms
correction to the Keck data, putting the two spectrometers'
measurements on the same velocity scale before showing them in Figure
\ref{fig1} and listing them in Table 1.  The first 14 Doppler
measurements made between 1989 and November 1994 typically have
uncertainties of 8--10 \mse, worse than most of the subsequent
observations due to the unrepaired optics of the Hamilton
spectrometer.  Observations made at Lick since December 1994 have
typical uncertainties of 3--5 \mse.  At Keck, between 1999 and 2004,
the typical Doppler uncertainty is 3 \mse.  In August 2004 the optics
and CCD detector for HIRES were upgraded, reducing the Doppler errors.
In November 2004 we began making three consecutive observations (and
sometimes five) of 55 Cnc to average over stellar p-mode
oscillations that can add 1 \ms velocity noise to G8V main sequence 
stars \citep{Kjeldsen05}.  The resulting Doppler precision at Keck since August
2004 has been 1.0--1.5 \mse.

\section{Keplerian Fits to Doppler Measurements}
\label{sec_kepfits}

The Doppler measurements of 55 Cnc were fit with a series of
Keplerian models, each model having an increasing number 
of planets, beginning with the two well established
periods of 14.65 d and 14.7 yr \citep{Marcy_55cnc}.  We polished all
models with a Marquardt minimization of \chisq to establish the
best-fit model.  The weights assigned to each Doppler measurement are
the inverse square of each measurement's uncertainty, which are
approximated as the quadrature sum of the internal uncertainty in the
Doppler measurement and the ``jitter'' that stems from photospheric
motions and instrumental errors \citep{Wright05}.  Experience with
similar G8 main-sequence stars suggests that the combined
astrophysical and instrumental jitter is 3 \ms at Lick and 1.5 \ms at
Keck, both values being uncertain by 50 percent. The jitter prediction
is complicated by the high metallicity of 55 Cnc, [Fe/H] = +0.3.  The
radiative transfer of \caii in 55 Cnc will be different from that in
solar metallicity stars, because of higher line and continuous
opacities, rendering the calibration of emission with stellar age, rotation,
and jitter even more uncertain.  However, the estimated rotation
period of 39 d from periodicities in the \caii emission 
and the star's low rotational \vsini of 2.5 \kms confirm that
the star has, at most, a modest level of magnetic activity, indicating
correspondingly modest jitter with an upper limit of 4 \mse.  For this
analysis, we adopt a jitter value of 1.5 \ms and 3.0 \ms for Keck and
Lick, respectively.

After fitting a model with the two well established planets, we
assessed the statistical significance of any periodicities remaining
in the residuals to motivate addition of another planet to the model,
as described in detail below.  We determine false alarm probabilities
for peaks in the periodogram attributed to any additional planets by
testing the null hypothesis that the current velocity residuals are
merely incoherent noise.  In such tests, the velocity residuals to our
best-fit model are scrambled and their periodograms computed to assess
the fraction of trials with scrambled residuals that have a stronger
peaks.  This FAP assessment makes few assumptions about the width or
shape of the distribution of noise.

\subsection{The Three-Planet Model}

Our initial model consisted of the sum of two Keplerian orbits (no
gravitational interactions) for the two planets
having secure orbital periods of 14.65 d and $\sim$14.7 yr, both
strongly supported by all of our past Doppler analyses of this star
\citep{Marcy_55cnc}.  A two-planet fit yields periods of 14.65 d and
14.7 yr and eccentricities of 0.002 and 0.06 for the two planets,
respectively.  The residuals have an RMS of 11.28 \ms and \cs of 3.42.
An accurate model would have an RMS on the order of the errors in the
data plus ``jitter'' ($\sim$5 \ms) and \cs near 1.  These large values
of RMS and \cs
indicate that the model is inadequate.  The periodogram of the
residuals (Figure \ref{fig2}) exhibits a tall peak at a period of 44.3
d and power of 55, clearly significant above the noise.  This period
corresponds to the orbit of the planet suspected in
\citet{Marcy_55cnc}.  This 44.3 d period is most likely caused by a
third planet as the only other explanation would be rotational
variation from surface inhomogeneities.  Such rotational explanations
are ruled out both by the shorter stellar rotational period, 42.7 d,
found in the photometry as shown in \S 6, and by the large velocity
amplitude of 10.6 \mse, which is never
seen in such chromospherically quiet stars. Furthermore, photospheric
features generally only survive for a few rotation periods of the
star.  It seems unlikely that surface inhomogeneities would persist
for more than a decade and maintain rotational phase coherence.

A Levenberg-Marquardt minimization was used to find the best-fit
orbital parameters for a three-planet Keplerian model with periods
near 14.65 d, 44.34 d, and 14.7 yr. The best fit yielded residuals
with an RMS scatter of 8.62 \mse and \cs = 2.50.  This result
represents an improved fit to the two-planet model, but is still
clearly inadequate, not surprising as the model did not include a
periodicity near 2.8 d as found by \citet{McArthur04}.  Indeed, the
periodogram of the residuals to the three-planet fit, shown in Figure
\ref{fig3}, reveals two additional strong peaks near 2.8 d and 260 d.

\subsection{The Four-Planet Model}

We proceeded to test a 4-planet model by including a fourth planet with a
period near 2.8 d \citep{McArthur04}.  The best-fit 4-planet model
gave periods of 2.81 d, 14.65184 d, 44.32 d, 14.4 yr, all with
eccentricities less than 0.3.  The residuals have RMS of 7.87 \ms and
\cs = 2.12, both representing a significant improvement over the
3-planet model.  (In computing both the RMS and \chisqe, the
denominator was appropriately diminished by the greater number of free
parameters, i.e., five per planet.)  Thus both the periodogram in
Figure \ref{fig3} and the superior fit with four planets offer support
for the existence of the planet with 2.81 d, corresponding to the
planet with $P$ = 2.808 $\pm$ 0.002 d in \citet{McArthur04}.

However this 4-planet model remains inadequate for two reasons.  The
residuals reveal a poor fit with \cs = 2.12 and an RMS of 7.87 \mse,
larger than explainable by the Doppler errors and jitter.  
Also, a periodogram of the residuals reveals a peak at a
period of 260.1 d, as shown in Figure \ref{fig4}, and some additional
smaller peaks.

\subsection{Assessing the Periodicity near 260 d}
\label{sec_260d}

The periodogram peak near 260 d (Figure 4) in the residuals to the
4-planet model could be spurious, caused by fluctuations arising from
photon-limited Doppler errors in the spectra or by aliases in the
window function of the sampling times.  The CCD detector at Lick
Observatory has been upgraded four times in the past eighteen years,
which could produce discontinuities of 1--2 \ms in their zero points
and even create an alias.  Such abrupt, one-time instrumental changes should not
produce periodicities.  Nonetheless, to check for such effects, the four-planet
Keplerian model was fit separately to the Lick and Keck velocities.

The 250 velocities from Lick against only 70 from Keck cause the
periodogram in Figure 4 to be heavily weighted toward the Lick
measurements.  The prominent period at 260 d certainly reflects the
Lick velocities more than those from Keck, leaving open the question
of independent confirmation of the 260 d period in the Keck data. We
fit a 4-planet model to the 70 Keck velocities alone.  The Keck
velocities offer higher precision ($\sim$1.5 \ms) than those from Lick
but carry the disadvantage of a duration of only 5 1/2 years.

The four-planet fit to the Keck velocities alone yielded residuals
with RMS = 4.3 \ms and \cs = 2.59.  The periodogram of the residuals
is shown in Figure \ref{fig5}, and it reveals a peak at a period of 266 d
with a power of 7.7.  There is no significant power at any other
periods.  The power in the 266 d peak is higher than all peaks for
periods between 1-3000 d.  Importantly, the peak at 266 d is roughly
twice as high as the noise peaks.  Although this peak is not
overwhelming by itself, the independent occurrence of a periodicity
near 265 d in the Keck velocities along with the similar period found
in the Lick velocities, supports the reality of that period and argues
against systematic errors as the cause.

One might be concerned that the Keck velocities yielded such a modest
peak at $\sim$266 d (Figure \ref{fig5}) as compared to the relatively
strong peak in the Lick data (Figure \ref{fig4}).  We addressed this
concern by augmenting the Keck velocities with artificial velocities
corresponding to a planet having a period of 260 d in a Keplerian
orbit that causes a semiamplitude of $K = 4.4$ \mse. The idea is 
that if the power in the periodogram doubles, then the modest peak 
in Figure 5 is probably reasonable. We performed a
four-planet fit and computed the periodogram of the residuals.  A peak
at $P = 261$ d was seen with a power of 13, roughly double the power
of the peak that emerged from the original velocities. Thus, the 266 d peak 
in the periodogram from the original Keck velocities (Figure 5) constitutes a
confirmation of the $\sim$265 d planet seen in Figure 4 having that
period and amplitude.  Of course, the Lick velocities alone also
exhibit the 260 d peak indepedently.

We also checked to see if the 260 d signal might be an alias of
the possible 470 d peak seen in Figure 4.  We fit the combined
Lick and Keck velocities with a five-planet model having a fifth planet
with a period near 470 d instead of near 260 d.  The periodogram of
the residuals to this five-planet model still has a strong peak with
period near 263 d, with a power of 19.  Apparently the period at
260 d does not vanish by including a 470 d period in the model
and thus is not an alias of it.

We assessed the probability that the 260 d signal was caused by
chance fluctuations in the velocities by performing a conservative
false alarm probability test. We fit the combined velocities with only
a four-planet model and tested the null hypothesis that no periodicity
near 260 d actually exists in the residuals, implying that the
peak is due merely to noise.  We scrambled the residuals to the
four-planet fit but kept the times of observation the same, and
recomputed the periodogram for each of 500 trials.  We recorded the
power of the tallest peak in the periodogram from each of 500 trials.
The histogram of those peak powers is shown in Figure \ref{fig6}.  The
typical peaks from the scrambled residuals have powers of 7--13, with
the tallest being 16.  In contrast, the periodogram of the original
residuals had a peak height of 31.5, shown both in \ref{fig4} and as
the vertical dashed line in \ref{fig6}.  Thus, the null hypothesis
(that the residuals have no coherence) is unlikely and the associated
false alarm probability of the peak at 260 d is less than 0.002,
indicating that the periodicity is real.

The analysis above strongly supports the existence of a planet with a
period of 260 d.  The period of 260 d does not correspond to any known
time scale of stellar interiors or atmospheres, nor to the rotation
period of the star which is 42.7 d (see below).  Thus, a plausible 
interpretation is a planet with a period near $P = 260$ d, making it
the fifth planet in the 55 Cnc system.

\subsection{The Five-Planet Model with a 260-day Planet}

We constructed a Keplerian model that included a fifth planet having a
period near 260 d.  A best-fit model to the combined Lick and Keck
velocities was found easily, yielding five periods of 14.65162 d,
44.344 d, 5218 d, 2.817 d, and 260.0 d (see Table 2).  The residuals
have RMS = 6.74 \ms and \cs = 1.67 (including the jitter in the
expected variance), and a periodogram of them is shown in Figure 7.  
The values of the RMS and \cs are 15\% and 20\%
lower, respectively, than the corresponding diagnostics of the
four-planet model. Table 3 gives the RMS and \cs for all multi-planet
models considered in this paper, showing the significant improvement
with each additional planet.  This major improvement in the quality of
the fit of 320 measurements, coming from a fifth planet with its five
additional free parameters, indicates that the new model has
considerable merit.  The five-planet model containing the 260 d
planet is clearly superior to the four-planet model.
The period agrees with
that found by \citet{Wisdom05a} from a periodogram analysis of our
earlier, published velocities from Lick Observatory.

As this model contains a proposed planet with $P = 260$ d, we
present in Table 2 all of the orbital parameters for all five planets
self-consistently computed with a Levenberg-Marquardt least squares
algorithm.  Considerable trial and error with various starting guesses
for the 26 free parameters was carried out to ensure that the least
squares search began near the deepest minimum.    
The \chisq fit was virtually unchanged for orbital eccentricites between 
0.0 to 0.4 for the 260 d planet.  This is not surprising since the 
amplitude of the planet is comparable to the single measurement 
precision for most of our data.  Although the orbit is consistent
with circular, we adopted an intermediate eccentricity of $0.2 \pm 0.2$ 
to indicate the indistinguishable range of eccentricity. The best-fit
parameters for the 260 d planet are $e$ = 0.2 $\pm$ 0.2, $K$ = 4.879 $\pm$
0.6 \mse, implying \msini = 0.144 $\pm$ 0.04 \mjupe.

The innermost planet has $P$ = 2.81705 $\pm$ 0.0001 d, $e$ = 0.07
$\pm$ 0.06, $K$ = 5.07 $\pm$ 0.53 \mse, and \msini = 0.034 \mjup =
10.8 \mearthe.  In comparison, \citet{McArthur04} found the inner
planet to have a period, $P$ = 2.808 $\pm$ 0.002 d, $e$ = 0.174 $\pm$
0.127, $K$ = 6.67 \mse, and \msini = 0.045 \mjup = 14.2 \mearthe.

There is no question that the planet with $P$=2.817 d is the planet previously identified
as having a period of 2.808 d \citep{McArthur04}.  The new minimum mass \msini =
10.8 \mearth is lower than the 14.2 \mearth previously reported in
McArthur et al..  These differences are not surprising as some of the
excess velocity variation previously left to be absorbed by the four
known planets is now accounted for by the fifth planet.

The phase-folded velocities for the 260-day planet are shown in Figure
8 after subtracting the sum of the computed velocities of the other
four planets from the measured velocities. The orbital eccentricity has 
been fixed to 0.2. The resulting residual velocities 
are plotted versus orbital phase and shown in Figure 8.  The residuals
reveal the 260 d period that had been detected in the periodogram and
the Keplerian model is overplotted.  The scatter has an RMS of 6.74 \mse.  The
error bars shown in Figure \ref{fig8} include the quadrature sum of
the internal errors (typically 2 \ms for Keck and 4 \ms for Lick) and
the ``jitter'' (1.5 \ms for Keck and 3 \ms for Lick).  Thus,
the scatter of 6.74 \ms is only slightly larger than the 
known internal errors and expected jitter (astrophysical and
instrumental).

\section{Residual Planets} 

Several explanations for the modest 6.74 \ms scatter in the residuals
are possible.  Perhaps we are underestimating our internal errors.
Perhaps the jitter for this metal-rich star is somewhat higher than
the average for G8 main-sequence stars.  Or perhaps there are other
planets that cause a sufficiently low signal that they are not apparent
in the periodograms but nonetheless add a few \ms of ``noise'' to the
velocities.

We assessed the detectability of a hypothetical 6th planet by adding
the velocities that would be induced by it to the observed velocities.
We fit a 5-planet model to these augmented velocities, allowing all 26
parameters to float.  We searched the periodogram of the residuals for
peaks that loom above those arising in the 5-planet fit of the actual
velocities (Figure 7).  Such peaks would have been identified as a candidate 6th
planet.  We considered orbital periods from 300 d - 4000 d and
determined the minimum \msini that produced a peak 50\% above any of
the peaks in the actual periodogram (i.e. above a power of 15).

The minimum detectable mass of a hypothetical 6th planet is a
sensitive function of its period and phase as those parameters 
determine how easily the signal can be absorbed
in the 5-planet model, avoiding detection.
Neighboring periods differing by a mere
few percent can produce periodogram peaks differing by a factor of two
simply due to differing commensurability with the 5 existing
planets.   A 6th planet in a mean motion resonance is particularly
capable of avoiding detection in the face of the five existing
planets.  Such fine structure aside, the simulations can be
characterized as follows.  For orbital periods of 300 - 850 d,
a 6th planet with Msini below 50 \mearth
would have eluded detection as the periodogram peaks
would not have loomed even 50\% above the noise.   For periods 850 d
- 1500 d, a 6th
planet could avoid detection by having \msini below 100 \mearthe.
For periods 1750 - 4000 d, planets below 250 \mearth would elude
detection. Thus such planets could exist around 55 Cnc and yet have
avoided
detection by our current 18 years of Doppler measurements.  Indeed
several such planets could exist in the large gap between periods of
260 d and 13 yr and probably maintain dynamical stability.

\section{Dynamical Simulations of the Multi-Planet System}

The models in this paper are based on the approximation that the
planetary orbits are Keplerian ellipses. In actuality, the radial
velocity variation of the parent star over nearly two decades of
observation is also affected by the mutual gravitational perturbations
between the planets.  As a concrete example, one can interpret the
5-planet fit in Table 2 as describing a set of osculating orbital
elements at the epoch JD 2447578.730 of the first radial velocity
observation.  By making a choice of epoch, one creates a unique
initial condition for a six-body integration of Newton's equations of
motion. When this integration is carried out, one finds radial velocity
deviations of $\Delta V >$ 25 \ms in comparing Keplerian and
Newtonian models at epochs near JD 2454000.

These deviations arise primarily from the orbital precessions of
planets b, c, and d that occur in the Newtonian model that are absent
from the Keplerian model.  Because the orbits are nearly circular,
a Keplerian 5-planet fit can, however, compensate for nearly all
of the precession through small adjustments to the orbital periods.

It is likely that one can obtain an improved chi-square by adopting a
self-consistent N-body model for the stellar reflex velocity (e.g.
Rivera et al. 2005). In addition to lowering the RMS of the fit, a
definitive model of this type allows for the correct characterization
of the possible 3:1 resonant relationship between planets b and c, and
can therefore give important clues to the formation history of the
system.  Adopting the Keplerian fit in Table 2 as an intial guess, we
used Levenberg-Marquardt minimization to obtain a self-consistent
5-planet dynamical fit to the radial velocity data sets. The resulting
orbital parameters of our dynamical fit are all quite similar to their
corresponding values in the Keplerian model, and are listed in Table
4. Our dynamical fit has $\sqrt{\chi^2}$=2.012 (without including any
jitter), and RMS=7.712 \ms. A more computationally expensive search
should be able to find orbital parameters that provide a slight
improvement to these values. We leave such an analysis to future work.

In order to assess the dynamical stability of our five-planet model,
we adopt the self-consistent orbital elements in Table 4 and integrate
the system forward for one million years from epoch JD 2447578.730. We
used a Bulirsch-Stoer integrator (Press et al. 1992). The system
remained stable throughout a one million year integration. The
evolution of the five planetary eccentricities during a representative
$2.5\times10^{4}$ year interval are shown in Figure \ref{fig10}. As is
true throughout the $10^6$ year integration, the eccentricity
variations experienced by all five planets are quite modest, and the
system appears likely to be dynamically stable for long periods.

It is interesting to note that during the course of the numerical
integration, the 3:1 resonant arguments for planets b and c are all
circulating. This indicates that planets "b" and "c" do not currently
participate in a low-order mean motion resonance, despite the near
commensurability of their orbital periods.

We have computed the eccentricity variations that result when the
system is modeled using a secular perturbation theory that includes
terms up to second order in eccentricity (see, e.g., Murray \& Dermott
1999), and which includes the leading-order effects of general
relativity as outlined by Adams \& Laughlin (2006).  The results are
quite similar to those in Figure \ref{fig10}.  This indicates that the
bulk of the planet-planet interactions in the system can be accounted
for with a simple second-order theory, thus improving the likelihood
that the configuration of planets can remain dynamically stable for
the lifetime of the star.

\section{Photometry of 55 Cancri}

We have used the T8 0.8~m automatic photometric telescope (APT) at
Fairborn Observatory to obtain high-precision photometry of 55 Cnc
during 11 observing seasons between 1996 November and 2007 April.  The
T8 APT is one of several automated telescopes at Fairborn dedicated to
observing long-term, low-amplitude brightness variations in solar-type
stars associated with stellar magnetic cycles as well as to measuring
short-term, low-amplitude variations caused by rotational modulation
in the visibility of surface magnetic features \citep{h99}.  APT
photometry of planetary candidate stars helps to establish whether
observed radial velocity variations are caused by stellar activity or
planetary-reflex motion, and direct measures
of stellar rotation periods provide good age estimates of the
planetary systems \citep[e.g.,][]{hetal00a}.  \citet{qetal01} and
\citet{petal04} have published several examples of periodic radial
velocity variations in solar-type stars caused by photospheric spots
and plages.  The APT observations are also useful to search for
transits of the planetary companions.  The rare transiting systems
allow direct determination of basic planetary parameters such as mass,
radius, and mean density and so provide observational constraints on
models of planetary composition and internal structure
\citep[e.g.,][]{setal05}.  Bright transiting systems enable detailed
follow-up studies of exoplanet atmospheres \citep[e.g.,][]{retal07}.
Finally, monitoring a planetary host star's long-term luminosity
variations provides a measure of the star's climate forcing ability
on its system of planets \citep[e.g.,][]{hhl07}.

The T8 APT is equipped with a
two-channel precision photometer employing two EMI 9124QB bi-alkali
photomultiplier tubes to make simultaneous measurements in the
Str\"omgren $b$ and $y$ pass bands.  The APT measures the difference
in brightness between a program star and one or more nearby constant
comparison star(s); the primary comparison star used for 55~Cnc is
HD~76572 ($V=6.28$, $B-V = 0.42$, F6~IV-V).  The Str\"omgren $b$ and
$y$ differential magnitudes are corrected for differential
extinction with nightly extinction coefficients and transformed to
the Str\"omgren system with yearly mean transformation coefficients.
Finally, the $\Delta b$ and $\Delta y$ observations are combined
into a single $\Delta(b+y)/2$ pass band to increase the photometric
precision.  The external precision of a single differential
magnitude is typically around 0.0015 mag for the T8 APT, as
determined from pairs of constant stars.  Further details on the
telescope, photometer, observing procedures, and data reduction
techniques can be found in \citet{h99}.  

The complete 11-yr set of differential magnitudes computed with the
primary comparison star is plotted in the top panel of Figure~11.
Intercomparison of the primary comparison star with two secondary
comparisons (HD~77190, $V=6.07$, $B-V = 0.24$, A8Vn; HD~79929,
$V=6.77$, $B-V = 0.41$, F6V) revealed that the annual means of the
primary comparison vary over a range of 0.003 mag from year to year.
Rather than switch to one of the more stable secondary comparison
stars, we have instead normalized the 11 seasons with the primary
comparison so they all have the same annual mean.  This was done
because the secondary comparison stars have been used only
for the
past seven years.  The normalization removes any long-term variation
in the primary comparison star as well as in 55~Cnc, but this
improves the sensitivity of our transit search described below for
orbital periods under one year.  After normalization, outliers
exceeding three standard deviations were removed.  The final data
set in the top panel of Figure~11 contains 1349 nightly
observations; the standard deviation of an individual observation
from the normalized mean is 0.0017 mag.  

The 0.0017 mag standard deviation of the full data set is only
slightly greater than the nominal measurement precision of 0.0015
mag but suggests that low-amplitude, short-term intrinsic
variability might be present at times in 55~Cnc.  (Long-term
variability has been removed by the normalization).  We searched
each annual set of measurements for evidence of coherent,
low-amplitude variability that might be the result of rotational
modulation in the visibility of starspots.  The middle panel of
Figure~11, which shows photometry from a portion of the 9th
observing season, exhibits the clearest example of coherent
variability in the data set. Two cycles of brightness variation are
visible with an amplitude of approximately 0.006 mag.  We interpret
this as evidence for a small starspot region (covering less than 1\%
of the star's visible surface) that has survived for two rotation
cycles of the star.  A power spectrum of the observations in the
middle panel is computed with the method of \citet{v71} and shown in
the bottom panel of Figure~11.  This gives a period of 42.7 $\pm$
2.5 days, which we interpret to be the stellar rotation period.
This confirms the rotation period of 55~Cnc reported by
\citet{hetal00a}, who used rotational modulation of the Ca II H \& K
flux measured by the HK Project at Mount Wilson Observatory
\citep{betal98}.

We searched the photometric data for evidence of
transits of the four inner planets; the results are summarized in
Table~4 and plotted in Figure~12.  We first computed the
semi-amplitudes of the light curves (column~3) with least-squares
sine fits of the complete data set phased to the four shortest
radial velocity periods.  The resulting amplitudes are all extremely
small and consistent with zero.  These very tight limits on
photometric variability on the radial velocity periods clearly
support planetary-reflex motion as the cause of the radial velocity
variations.  While our measured 42.7 day rotation period is
consistent with the 44.35 day radial velocity period because of the
relatively large uncertainty of 2.5 days in the rotation period, the
absence of any photometric variability on the more accurate 44.35
day radial velocity period is strong support for the existence of
55~Cnc c.  

In Figure~12, we have plotted light curves of the
photometric data phased with the orbital periods of the inner four
planetary companions.  Zero phase in each panel represents the
predicted phase of mid transit for each of the companions.  Only
phases from 0.94 to 0.06 are plotted to improve visibility of any
possible transits.  The solid curve in each panel approximates the
predicted transit light curve, assuming a planetary orbital
inclination of 90\arcdeg (central transits).  The out-of-transit
light level corresponds to the mean (normalized) brightness of the
observations.  The transit durations are calculated from the orbital
elements, while the transit depths are derived from the estimated
stellar radii and the planetary radii computed with the models of
\citet{bll03}.  The horizontal error bar below each predicted
transit curve represents the approximate $\pm 1 \sigma$ uncertainty
in the time of mid transit, based on Monte Carlo simulations and the
uncertainties in the orbital elements.  The vertical error bar
represents the typical $\pm 0.0015$ mag uncertainty of a single
observation.  

Column~4 of Table~5 lists the geometric probability of
transits for each of the five companions, computed from equation 1 of
\citet{setal03} and assuming random orbital inclinations.  The
predicted transit depths for each planet determined as described
above are given in Column~5, and the ``observed transit depths'' are
recorded in column~6.  The observed depths are computed as the
difference in the mean light levels between observations that fall
inside and outside of the transit windows plotted in Figure~12; a
positive depth indicates a brightness drop in the transit
window.  

Unfortunately, we see no evidence for transits of the inner,
low-mass companion 55~Cnc~e; the mean of the 51 observations within
its predicted transit window agree with the out-of-transit
observations within 0.00029$\pm$0.0002 mag, a result consistent with
the absence of 0.00065 mag transits but still allowing a small
possibility for their existence.  Since the uncertainty in the time
of mid transit is rather large compared to the transit duration, we
searched for shallow transits over the full range of phases and over
orbital periods between 2.70--2.90 days with null results.  The
secure detection of transits of such a small body would require
reducing the uncertainty of the in-transit brightness mean by a
factor of about two, which would require a factor of four more
observations.  This could be accomplished with the APT over the next
observing season or two by concentrating brightness measurements
around the times of predicted transits.  We are forced to leave our
non-detection of transits of 55~Cnc~e as an uncertain result, as
indicated by the colon in column~7 of Table~5.

Table~5 and Figure~12 demonstrate that transits with the expected depths of
55~Cnc~b and 55~Cnc~c do not occur.  The observed transit  depths are
both consistent with zero.  Figure~12 shows that we have no
photometric observations during the predicted time of transit
of 55~Cnc~f.  However, given the uncertainty in the transit timing
and the density of observations within the uncertainty range, we
conclude that transits of planet f probably do not occur.  We have
insufficient radial velocities to predict accurate transit times of
the outermost planet 55~Cnc~d, so we can say nothing about their
occurrence.  We note that our non-detection of transits is consistent
with the likely inclinations of the planetary orbits as discussed in
\S{7} (below).  

Finally, we comment on the long-term variability
of the host star 55~Cnc. Although our normalization of the light
curve has removed any such variation from the present analysis, an
examination of the light curves computed with the secondary
comparison stars mentioned above show that 55 Cnc clearly exhibits
year-to-year variations in mean brightness with an amplitude of 0.001
mag over a time scale of several years or more (Henry et al.,in
preparation).  Thus, long- and short-term brightness variations in
55~Cnc are very similar to irradiance variations in our Sun
\citep[.e.g.,][]{w77}.

\section{Minimum Mass Protoplanetary Nebula}

Planet formation in the protoplanetary disk around 55 Cnc was
apparently extraordinarily efficient, yielding at least five planets.
Our extensive radial velocity data set, with its 18-year baseline,
gives no indication that additional Jupiter-mass companions exist
beyond 6 AU, although Saturn-mass or smaller planets would easily go
undetected.  If, as a thought experiment, we grind up the currently
known planets, we may infer the properties of the protoplanetary disk
around 55 Cnc.

We first assume an edge-on, co-planar geometry with $i=90^{\circ}$, 
and {\it in situ} formation of the planets directly from the disk gas,
e.g. \citet{Boss97}.  In this case, the masses of planets total $\sim
5.3 M_{\rm Jup}$ and together imply a lower limit of 410 g cm$^{-2}$
for the average surface density of the disk interior to 6 AU. For a
gas-to-dust ratio of 100, this implies an average surface density in
solids of 4 g cm$^{-2}$. If the radial surface density profile of 55
Cancri's protostellar disk declined as $\sigma(r)\propto r^{-3/2}$,
this implies a solid surface density of 1.4 g cm$^{-2}$ at 5 AU, which
is approximately half the value of the minimum-mass solar nebula at
Jupiter's current position.

In all likelihood, however, the surface density of solids in 55
Cancri's protostellar disk was higher than in the solar nebula. If we
assume that the planets formed via the core accretion mechanism, as
described, for example, by \citet{Hubickyj05} we estimate that they
contain at least 150 \mearth of heavy elements. Here we include
the high metallicity of the host star, 55 Cnc, with its [Fe/H]=+0.3 as
representative of the planet's interior.  Reconstituting this mass of
solids to recover 55 Cancri's metallicity implies an original
protostellar disk mass of $\sim 0.025 M_{\odot}$ within 6 AU.
Assuming that the nascent disk extended to 30 AU with a $r^{-3/2}$
surface density profile, the total mass would have been $\sim 0.06
M_{\odot}$, and the surface mass density in solids at 5 AU would have
been 7 g cm$^{-2}$.  

Adopting the reported orbital inclination of 53 deg for the outer
planet \cite{McArthur04} and assuming the orbits to be co-planar
augments all masses by 1/$\sin i$ = 1.25.  The resulting simplistic
minimum mass protoplanetary nebula has 510 g cm$^{-2}$ for the average
surface mass density of the combined gas and dust within 6 AU.
Adopting a nominal gas to dust ratio of 100 yields a dust surface mass
density of 5 g cm$^{-2}$.

But again considering the likely enrichment of solids within giant
planets, and the associated H and He, yields an original mass within 6
AU of at least 0.031 \msune.  Extending this disk to 30 AU gives a
total mass of 0.075 \msune.  The estimated surface mass density of
solids in the disk at 5 AU would have been 8.7 g cm$^{-2}$.

For expected equilibrium disk temperatures, this minimum mass disk is
below the threshold required for the development of non-axisymmetric
gravitational instabilities \citep{Laughlin96}, but likely high enough
to support the formation of planets via core accretion
\citep{Robinson06}.  In the context of this disk-profile scenario, the
core accretion theory suggests that additional objects with masses
ranging from Neptune to Saturn mass should be present beyond the
frontier marked by the orbit of planet d, i.e. beyond 6 AU.
If the outer planet migrated during or after its formation, the
estimated disk properties computed here would be affected.

\section{Discussion}

Our eighteen year campaign of Doppler measurements of 55 Cnc at the
Lick and Keck Observatories has gradually revealed additional
superimposed wobbles, each best interpreted as due to another orbiting
planet.  The previously identified four planets revealed a large gap
between 0.24 and 5.8 AU raising questions about unseen planets there
and the planet formation history in the protoplanetary disk.  The
velocities presented here reveal a fifth periodicity with $P = 260$ d,
consistent with Keplerian motion for which the most reasonable
interpretation is another orbiting planet.  The five-planet model
suggests the new planet has a minimum mass of 45 \mearth in a nearly
circular orbit with a semimajor axis of $a = 0.781$AU.  Thus, 55 Cnc 
is the first quintuple-planet system known.

This fifth planet apparently resides in the previously identified  gap
between 0.24-5.8 AU, and it remains between 
0.73 AU (periastron) and 0.84 AU (apastron), preventing orbit
crossings with both the next inner planet, ``c'', whose apastron is at
0.26 AU and the outer planet, ``d'', whose periastron is at 5.5 AU,
ensuring dynamical stability that is demonstrated numerically by 
N-body simulations.  As the star's luminosity is $L = 0.60$
\lsun (from its effective temperature and radius), this fifth planet 
resides within the classical habitable zone.  With a minimum mass of 
45 \mearthe, we speculate that it contains a substantial
amount of hydrogen and helium, not unlike Saturn ($M = 95$ \mearthe) 
in the solar system.

The four previously published planets around 55 Cancri now have
revised orbital parameters and masses because the fifth planet had
been contaminating the Doppler signal but was not taken into account.
The orbital semimajor axes and masses of all five planets (moving
outward from the central star) are $a = 0.038$ AU and \msini = 10.8
\mearthe; 0.115 AU and 0.824 \mjupe; 0.24 AU and 0.169 \mjupe; 0.781
AU and 0.144 \mjupe; and 5.77 AU and 3.83 \mjupe.  All quoted minimum
masses are uncertain by $\sim$5\% due to the uncertain mass of the
host star 55 Cnc.  The planets in this system all have nearly 
circular orbits, with the caveat that the orbital eccentricity for 
the 260 d planet is poorly constrained by radial velocity data. 

The inclination of the orbital plane of the outer planet has been
estimated from the apparent astrometric motion of the star, as
measured with the Fine Guidance Sensor of the Hubble Space Telescope
\citep{McArthur04}.  The derived orbital inclination is $i = 53 \pm
6.8$ deg (37 deg from edge-on) for that outer planet, implying that
its actual mass is 4.9 \mjupe.  Assuming coplanarity for the other
four planets, their actual masses (proceeding outward) are $M_e =
13.5$ \mearth (nearly one Uranus mass), $M_b = 1.03$ \mjupe, $M_c =
0.21$ \mjup = 66.7 \mearthe, and $M_f = 0.18$ \mjup = 57
\mearthe. Normally, co-planarity should not be a foregone conclusion,
as the eccentricities of many exoplanets imply a dynamically
perturbative history.  But for the 55 Cnc system with its five planets
so vulnerable to instabilities, such an active past
seems unlikely.  Any great perturbations would have ejected the
smaller planets.  Thus the planetary orbits in 55 Cnc are likely to
reside in a flattened plane, analogous to the ecliptic, coinciding
with the original protoplanetary disk out of which the planets
presumably formed.

If 55 Cnc did have a quiescent past, the star's spin 
axis should be nearly coincident with the normal of the system's
orbital plane.  The inclination of the spin axis can be determined
from the spectroscopically measured rotational \vsini = 2.46 \kms and
the photometrically determined spin period of 42.7 d for the star, along
with its radius of 0.93 \rsune.  The 42.7 d spin period implies an
equatorial velocity of only 1.24 \kmse, which is lower than \vsinie, 
an impossibility.  Either the measured \vsini is too high by a
factor of two (quite possible given the many line broadening sources)
or the spin period is much shorter (not likely, given the low
chromospheric activity). As is common, the inclination of the spin
axis of a nearby star, as attempted here, carries uncertainties so
large as to be of little value.  Nonetheless the crude interpretation
would be that the star is not viewed pole-on.  Indeed, if the orbits
were viewed nearly pole-on the implied planetary masses would be so
large as to render the system dynamically unstable.  We conclude that
the orbital plane of 55 Cnc is not being viewed nearly pole-on,
consistent with the astrometric value of $i$=53 deg.

As the 55 Cnc system has more planets than any previously discovered
system, its overall structure, including its planet mass distribution,
its density of orbits, and its orbital eccentricities, offers direct
constraints about its protoplanetary disk and subsequent planetary
dynamics. 
The 55 Cnc system may initially be sketched as having one
massive planet, likely a hydrogen-helium gas giant, in a nearly
circular orbit at 5.8 AU.  Inward are four less massive
planets, the innermost being roughly Uranus mass, the next outward
having roughly Jupiter mass and likely gaseous, and the next two
having somewhat sub-Saturn masses, also probably gaseous.  The outer
planet has an orbital angular momentum quickly shown to be at least
8.2$\times$10$^{43}$ kg m$^2$ s$^{-1}$, certainly 100 times greater
than the star's {\it spin} angular momentum, 6.1$\times$10$^{41}$ kg
m$^2$ s$^{-1}$ (from its 42.7 d spin period).

Thus, 55 Cnc contains a dominant outer planet at 5.8 AU of roughly 5
\mjup and four lower mass planets, all five in nearly circular orbits
(though two orbits appear to be somewhat more eccentric than found for 
the more massive planets in our solar system).  The large eccentricities 
found in the majority of
exoplanets \citep{Marcy_Japan_05, Butler06} are not seen in 55 Cnc.
The five orbits in 55 Cnc are probably nearly co-planar, as discussed
above, lest some planet masses be too large to allow stability.  Thus,
55 Cnc system has some basic structural attributes found in our solar
system: nearly coplanar, circular orbits, with a dominant gas giant
between 5-6 AU.  This similarity suggests that such solar system
architectures are not extremely rare.

The formation of multi-planet systems with outer, dominant planets may
occasionally form such that they persist for billions of years without
disruptive gravitational perturbations that cause large eccentricities
and eject planets.  Because nested, coplanar, circular orbits could
hardly be obtained unless they began that way, the 55 Cnc system, along
with the solar system, supports the hypothesis that planets form in
viscous protoplanetary disks, as has long been predicted by standard
planet formation theory, e.g., \citet{Lissauer95}.

One puzzle is whether the 55 Cnc planets suffered significant
migration.  The 44.35 d and 14.65 d have a period ratio of
3.027:1.000, thus leaving open the possibility of a mean motion
resonance identified previously \citep{Marcy_55cnc}.  As planets have
no reason to form with integer period ratios, any resonance suggests
that some differential migration occurred, allowing the two planets to
capture each other.  However the 3:1 mean motion resonance was found 
to be absent in the current N-body model as none of the relevant 
resonant arguments are librating. 

However, if migration occurred, we wonder what prevented the outer
planet, and indeed all of the planets, from migrating inward.  Perhaps
the disk dissipated just as this last crop of five planets formed, as
suggested in some migrational models \citep{Lin00, Trilling02,
Armitage02, Armitage03, Ida04a, Ida05, Narayan05}.  Indeed, the
proximity of the three inner planets to the host star, especially the
Jupiter-mass planet at 0.115 AU ($P = 14.65$ d), suggests that they
migrated inward to their present locations, assuming they did not form
in situ.  If so, protoplanetary disk material likely orbited outside
0.24 AU, exerting an inward torque on those planets and carrying away
orbital angular momentum in the system. During the migration period,
the implied disk material would have had a mass comparable to (or
exceeding) the Jovian-mass planets, from which the most recently
identified planet at 0.78 AU could have formed.

It is interesting that the third and fourth planets (at 0.24 and
0.78 AU) have small minimum masses, under 0.2 \mjupe, but are
surrounded by much larger giant planets with minimum masses of 0.824
and 3.8 \mjupe.  One wonders why the acquisition of material was
apparently so different among these four planets.

One also wonders why this particular star ended up with five planets
while 90 percent of stars on Doppler surveys do not have any 
detected giant planets. A statistical analysis 
of the planet detectability and observational incompleteness has been 
carried out by Cumming \etal 2007. Perhaps the high metallicity
of 55 Cnc ([Fe/H] = +0.30) played a role in the efficient planet
formation. Fischer \& Valenti (2005) find a correlation not only 
between stellar metallicity and the occurrence of planets, but also 
between high metallicity and multi-planet systems.  But we doubt 
that a mere factor of two enhancement in
heavy elements could account entirely for the five planets in this
system.  Some stochasticity in planet formation and subsequent
stability must play a role.

The outer planet at 5.8 AU is angularly separated from the star
($d = 12.5$ pc}) by 0.47 arcseconds making it a good target for
next-generation adaptive optics systems. The Space Interferometry
Mission, ``SIM PlanetQuest'', operating in narrow angle mode with
astrometric precision of 1 $\mu$as could measure the astrometric wobble
caused by all four outer planets, providing definitive masses and
orbital inclinations for them.  A spaceborn coronagraph or a spaceborn
interferometer might be capable of imaging the outer planet and taking
spectra of it.  NASA and ESA have a wonderful opportunity to fund such
an imaging telescope, thereby detecting and spectroscopically
assessing a mature extrasolar planet.  Moreover, as 55 Cancri is
metal-rich, the planets may also be abundant in heavy elements,
offering an opportunity to study rich atmospheric chemistry, clouds,
and weather, if spectra could be obtained.  Transits, if any occur,
would provide planet radii offering information about potential rocky 
cores.  

This rich planetary system portends a fruitful future for the Doppler
technique of studying exoplanets.  It shows that extending the time
baseline of Doppler measurements can reveal multiple planets, 
the existence or absence of which provides information about the 
formation, structure, and evolution of planetary systems.

\acknowledgements 

We gratefully acknowledge
the efforts and dedication of the Lick and Keck Observatory staff. 
We thank Karl Stapelfeldt for helpful comments. We thank the 
anonymous referee for comments that improved the manuscript. We
appreciate support by NASA grant NAG5-75005 and by NSF grant
AST-0307493 (to SSV); support by NSF grant AST-9988087, by NASA grant
NAG5-12182 and travel support from the Carnegie Institution of
Washington (to RPB). GWH acknowledges support from NASA grant NCC5-511
and NSF grant HRD-9706268.  We are also grateful for support by Sun
Microsystems.  We thank the NASA and UC Telescope assignment
committees for allocations of telescope time toward the planet search
around M dwarfs.  This research has made use of the Simbad database,
operated at CDS, Strasbourg, France. The authors wish to extend
special thanks to those of Hawaiian ancestry on whose sacred mountain
of Mauna Kea we are privileged to be guests.  Without their generous
hospitality, the Keck observations presented herein would not have
been possible.


\clearpage

\begin{deluxetable}{rrrr}
\tablenum{1}
\tablecaption{Velocities for 55 Cancri: Lick \& Keck}
\label{}
\tablewidth{0pt}
\tablehead{
\colhead{JD}         & \colhead{Vel.}     & \colhead{Unc.} & \colhead{Tele.} \\
\colhead{-2440000}   & \colhead{(\ms)}  & \colhead{(\ms)} & \colhead{}  }
\startdata
  7578.730 &   25.67 &    9.7 &   L  \\
  7847.044 &    3.91 &    8.4 &   L  \\
  8017.688 &   31.45 &    7.5 &   L  \\
  8375.669 &  -31.38 &    8.8 &   L  \\
\enddata
\tablecomments{Table 1 is presented in its entirety in the electronic edition
of the Astrophysical Journal.  A portion is shown here for guidance regarding
its form and content.}
\end{deluxetable}

\clearpage

\thispagestyle{empty}
\begin{deluxetable}{rrrrrrrrrr}
\tablenum{2}
\rotate
\tablecaption{Orbital Parameters for the Five-Planet Model}
\label{candid}
\tablewidth{0pt}
\tablehead{
\colhead{Star}  & \colhead{Period} & \colhead{$T_p$}  &  \colhead{$e$} & \colhead{$\omega$} & \colhead{$K$}   & \colhead{M$\sin i$}   & \colhead{$a$}  \\
\colhead{ }     & \colhead{(days)} & \colhead{}       & \colhead{ }    & \colhead{(deg)}    & \colhead{(\mse)} & \colhead{(M$_{Jup}$)} & \colhead{(AU)}
}
\startdata  
\tablenotemark{1}
Planet e   &  2.81705   &  2449999.83643 &  0.07  & 248.9   & 5.07    & 0.034   & 0.038   \\
$\pm$      &  0.0001    &     0.0001     &  0.06  &  38     & 0.53    & 0.0036  & 1.0$\times10^{-6}$ \\
Planet b   &  14.65162  &  2450002.94749 &  0.014 & 131.94  &  71.32  & 0.824   & 0.115   \\ 
$\pm$      &   0.0007   &       1.2      &  0.008 &  30     &  0.41   & 0.007   & 1.1$\times10^{-6}$ \\
Planet c   &  44.3446   &  2449989.3385  &  0.086 &  77.9   & 10.18   & 0.169   & 0.240   \\
$\pm$      &   0.007    &     3.3        &  0.052 &  29     & 0.43    & 0.008   & 4.5$\times 10^{-5}$ \\
Planet f   &  260.00    & 2450080.9108   &  0.2(f) & 181.1  & 4.879   & 0.144   & 0.781   \\
$\pm$      &   1.1      &      1.1       &  0.2   &  60.    & 0.6     &  0.04   & 0.007    \\
Planet d   &  5218      &  2452500.6     &  0.025 & 181.3   & 46.85   & 3.835   & 5.77    \\
$\pm$      &   230      &     230        &  0.03  &  32     &  1.8    & 0.08    & 0.11   \\
\enddata
\tablenotetext{1}{Planets are listed in order of increasing orbital period, 
however the planet designations, b-f, correspond to the chronological order of their discovery.}  
\tablenotetext{f}{ecc. fixed}  
\end{deluxetable}

\clearpage

\begin{deluxetable}{ccc}
\tabletypesize{\scriptsize}
\tablenum{3}
\tablewidth{0pt}
\tablecaption{Summary of improvments in RMS and \chisq fits}
\tablehead{
\colhead{Planet} & \colhead{RMS [\mse]} & \colhead{\cse} }
\startdata
b,c          &  11.28   &   3.42  \\  
b,c,d        &   8.62   &   2.50  \\
b,c,d,e      &   7.87   &   2.12  \\
b,c,d,e,f    &   6.74   &   1.67  \\
\enddata
\end{deluxetable}

\clearpage

\begin{deluxetable}{rlllllllll}
\tablenum{4}
\rotate
\tablecaption{Orbital Parameters from Self-Consistent Dynamical fit}
\label{selfconfit}
\tablewidth{0pt}
\tablehead{
\colhead{Planet}  & \colhead{Period} & \colhead{$T_p$}  &  \colhead{$e$} & \colhead{$\omega$} & \colhead{$K$}   & \colhead{M$\sin i$}   & \colhead{$a$} \\
\colhead{} & \colhead{(days)} & \colhead{(JD-2440000)}& \colhead{ } & \colhead{(deg)} & \colhead{(\mse)} & \colhead{(M$_{Jup}$)} & {(AU)}    
}
\startdata
55~Cancri b \tablenotemark{a} &  14.651262 & 7572.0307 & 0.0159  & 164.001 & 71.84  & 0.8358 & 0.115 \\
55~Cancri c  &  44.378710 & 7547.5250 & 0.0530  & 57.405  & 10.06  & 0.1691 & 0.241 \\
55~Cancri d  &  5371.8207 & 6862.3081 & 0.0633  & 162.658 & 47.20 & 3.9231 & 5.901 \\
55~Cancri e  &  2.796744  & 7578.2159 & 0.2637  & 156.500 & 3.73  & 0.0241 & 0.038 \\
55~Cancri f  &  260.6694  & 7488.0149 & 0.0002  & 205.566 & 4.75  & 0.1444 & 0.785 \\
\enddata
\tablenotetext{a}{Epoch= JD 2447578.730, $\sqrt{\chi^{2}}=2.012$
  (without jitter included), RMS=7.71 ${\rm m s^{-1}}$.  }
\end{deluxetable}

\clearpage

\begin{deluxetable}{ccccccc}
\tabletypesize{\scriptsize} 
\tablenum{5} 
\tablewidth{0pt}
\tablecaption{Results of Photometric Transit Search}
\tablehead{
\colhead{} & \colhead{Planetary} & \colhead{} & \colhead{Transit} & \colhead{Predicted} & \colhead{Observed} & \colhead{} \\
\colhead{} & \colhead{Period} & \colhead{Semi-Amplitude} & \colhead{Probability} & \colhead{Transit Depth} & \colhead{Transit Depth} & \colhead{} \\
\colhead{Planet} & \colhead{(days)} & \colhead{(mag)} & \colhead{(\%)}& \colhead{(mag)} & \colhead{(mag)} & \colhead{Transits}
}
\startdata 
e & 2.79565 & 0.00004 $\pm$ 0.00006 & 9.7 & $+$0.00065 & $+$0.00029$\pm$0.00020 & No? \\ 
b & 14.65165 & 0.00006 $\pm$ 0.00006 & 4.1 & $+$0.0143 & $+$0.0007$\pm$0.0005 & No \\ 
c & 44.3401 & 0.00008 $\pm$ 0.00006 & 2.0 & $+$0.0086 & $-$0.0003$\pm$0.0006 & No \\ 
f & 260.81 & 0.00008 $\pm$ 0.00006 & 0.8 & $+$0.0090 & \nodata\tablenotemark{a} & No: \\ 
d & 5223 & \nodata\tablenotemark{b} & 0.1 & $+$0.0155 & \nodata\tablenotemark{c} & ?  \\ 
\enddata
\tablenotetext{a}{No observations in predicted 12-hr transit window but many observations within the one sigma uncertainty interval.}
\tablenotetext{b}{Duration of the photometric record is less than the
planetary orbital period.} 
\tablenotetext{c}{Poorly constrained orbit and insufficient photometric phase coverage.}
\end{deluxetable}

\clearpage

\clearpage

\begin{figure}
\plotone{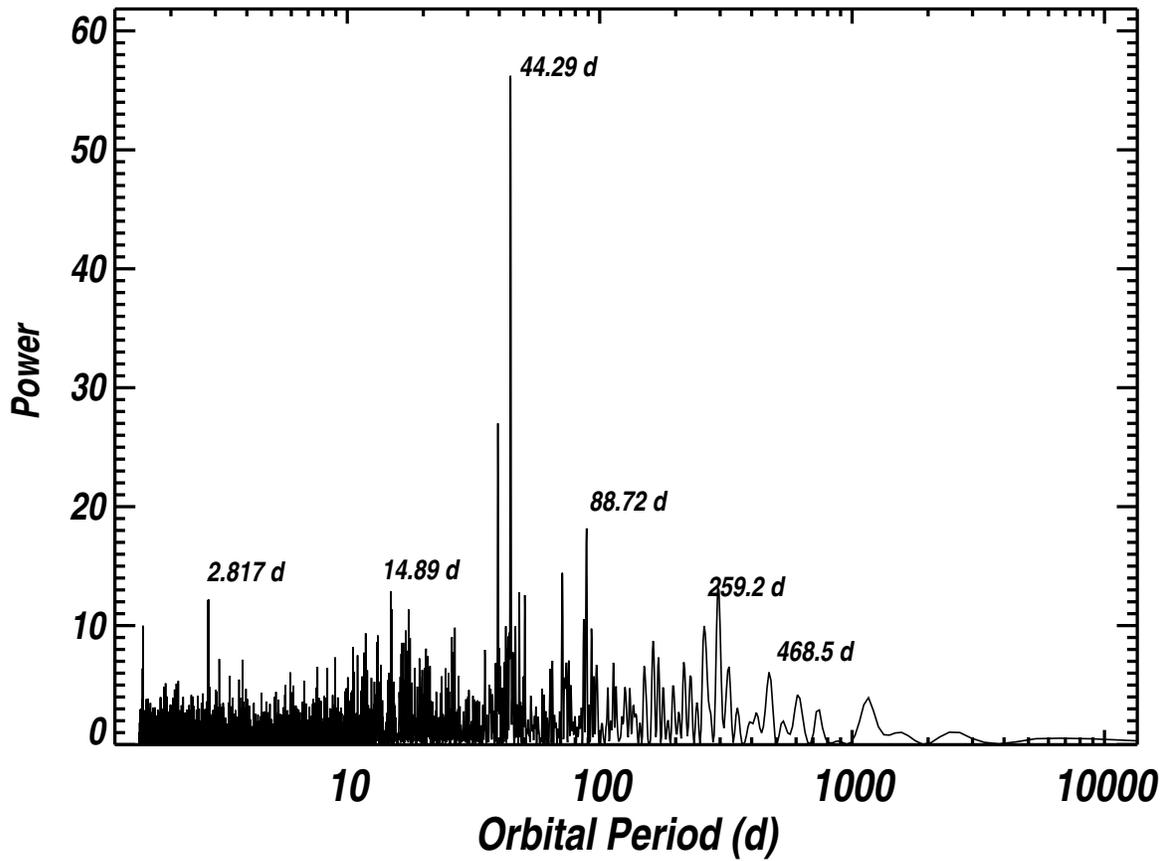}
\caption{Periodogram of the residuals to a Keplerian model that
  contains only the two well established planets with periods of 14.65 d and
  5200 d.  The tall peak at P = 44.3 d
  confirms the previously suspected planet with that period.}
\label{fig2}
\end{figure}
\clearpage

\begin{figure}
\plotone{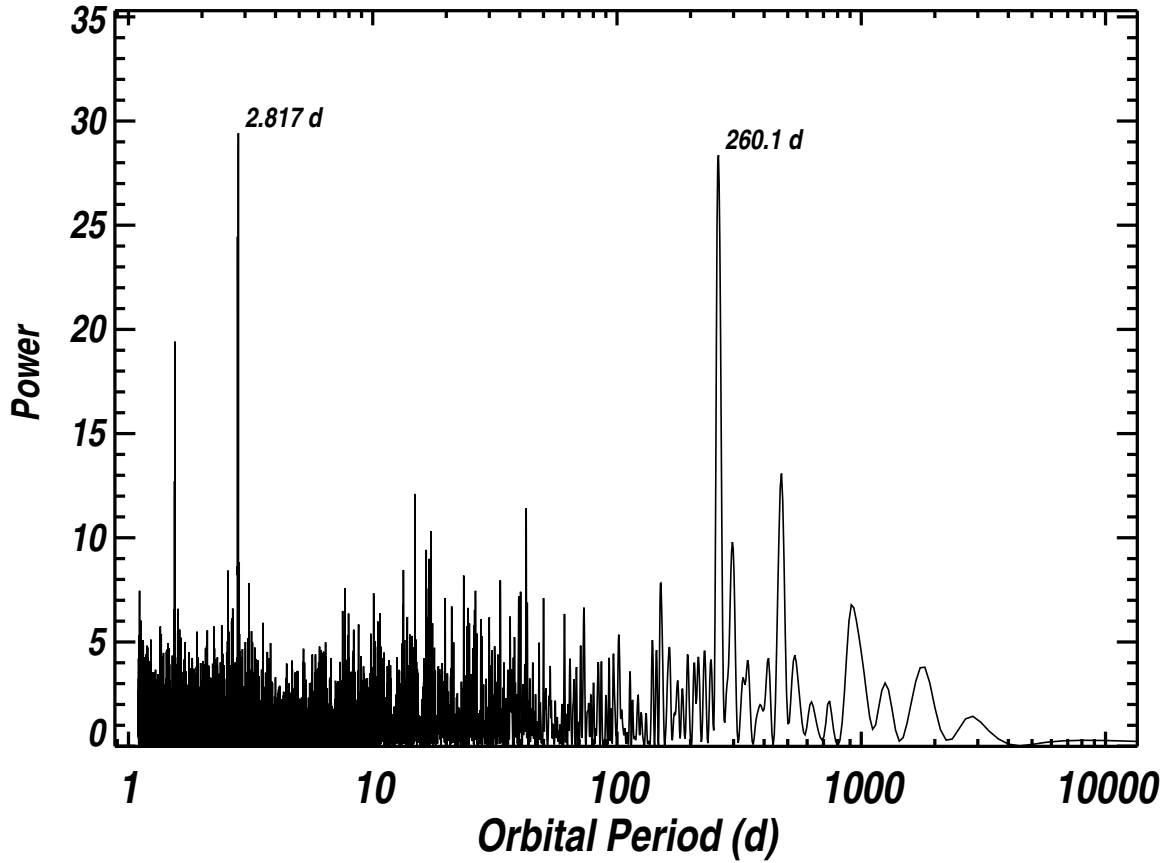}
\caption{Periodogram of the residuals to a Keplerian model that
  contains three known planets with periods of 14.6 d, 44.3 d, and
  5200 d.  The tallest peaks are at 2.81 d and 260 d suggesting the
  existence of real periodicities in the velocities.
  The peak at $\sim$1.5 d is an alias of the 2.8 d peak, and the peak at 460
  d is an alias of that at 260 d that disappears after modeling all five planets.}
\label{fig3}
\end{figure}
\clearpage

\begin{figure}
\plotone{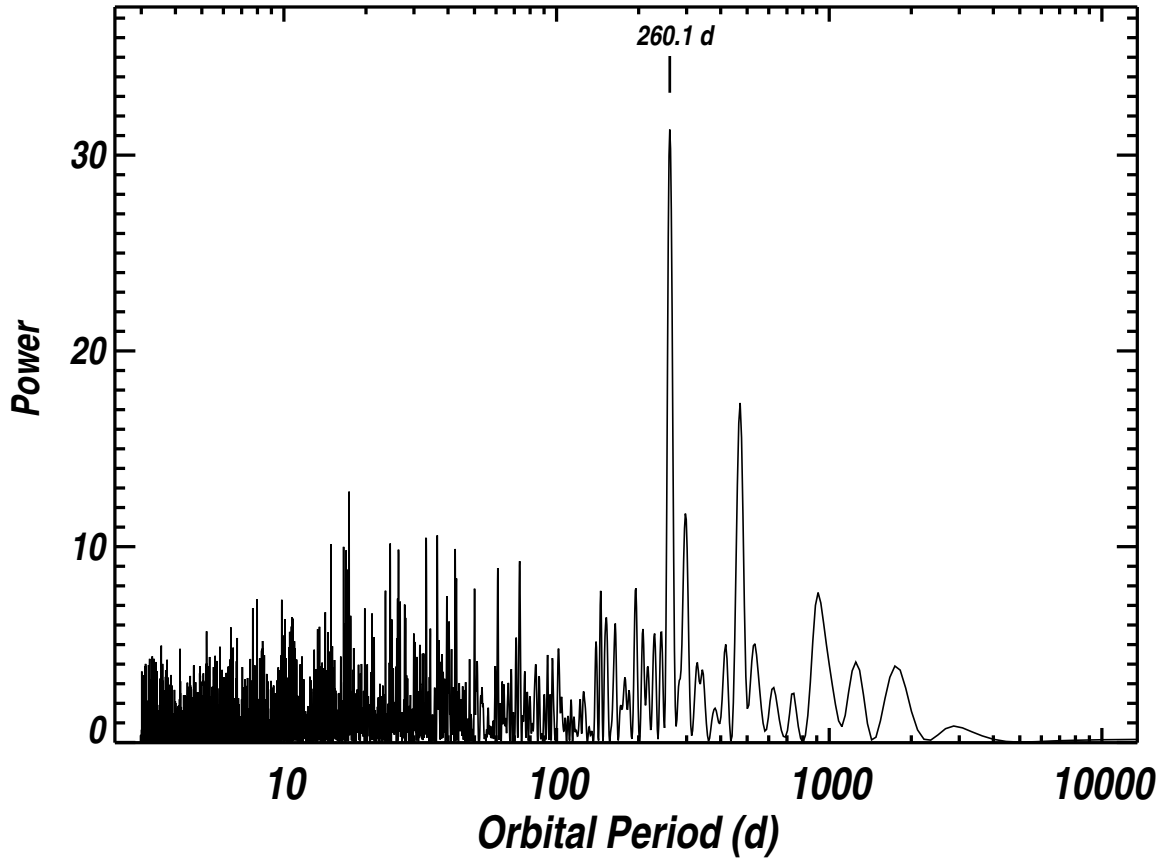}
\caption{Periodogram of the residuals to a Keplerian model that
  contains the four  previously suspected planets with periods near 2.817 d, 14.65
  d, 44.3 d, and 5200 d.  The periodogram exhibits a peak
  at 260.0 d, caused by the prospective fifth planet in the system.  
  The smaller peak to its right at 460 d is an alias.}
\label{fig4}
\end{figure}
\clearpage

\begin{figure}
\plotone{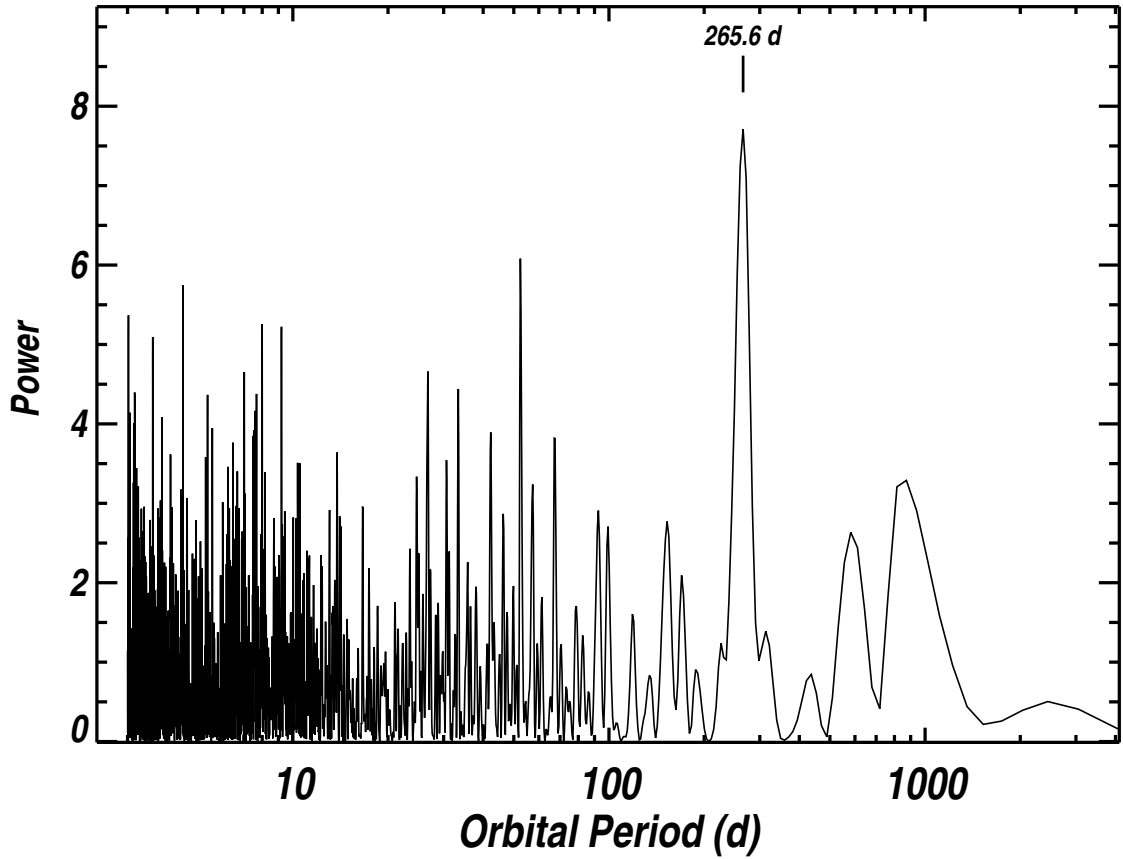}
\caption{Periodogram of the residuals to a 4-planet Keplerian model,
as in Figure 4, but fit to the Keck velocities only.  The tallest peak is
at a period of 265.6 d, nearly the same as that emerging from the Lick
data.  The modest peak power of only 7.7 is consistent with the
limited time sampling and duration of the Keck observations.  No other
period is compelling between periods of 1 and 3000 d.}
\label{fig5}
\end{figure}
\clearpage

\begin{figure}
\plotone{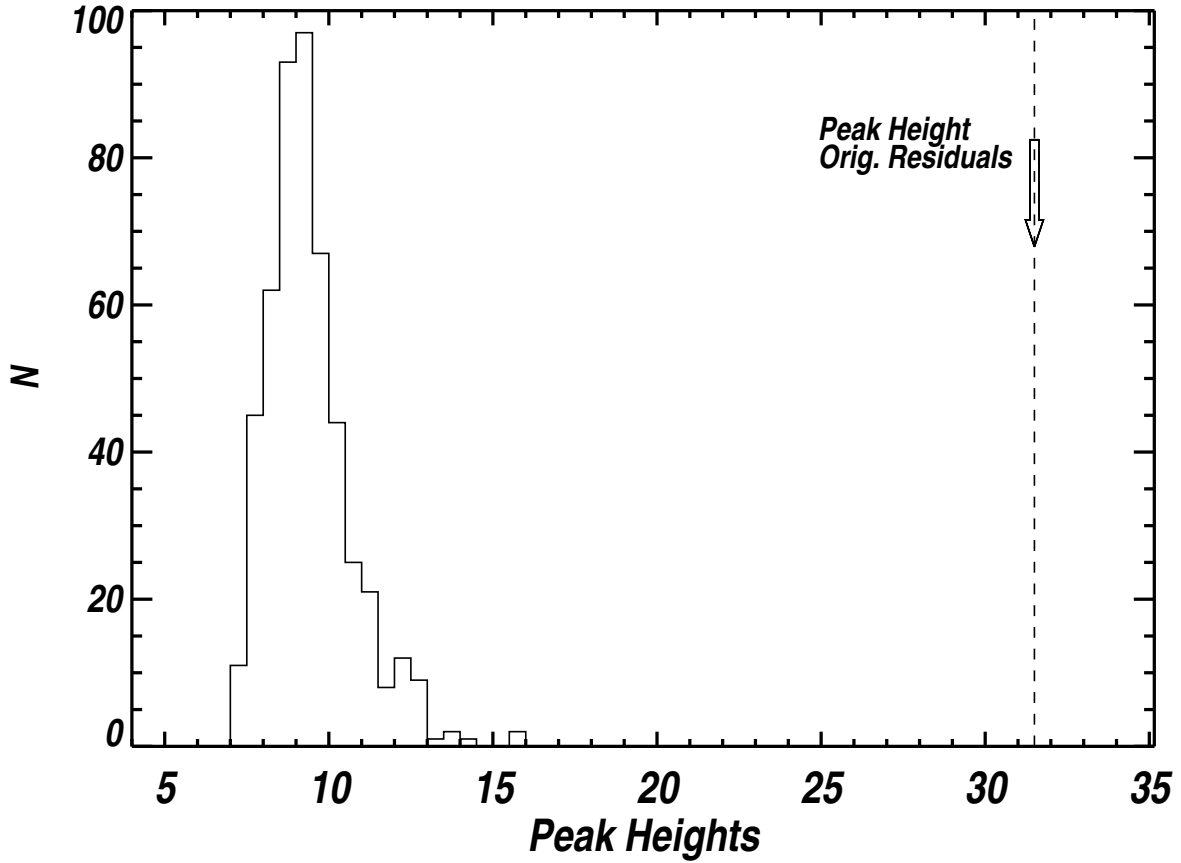}
\caption{A test of the False Alarm Probability of the 260 d periodicity
  seen in the combined velocities from both Lick and Keck (Figure 4).  
  The residuals to a 4-planet model were scrambled 500 times,
  yielding a histogram of the highest power in the periodograms.
  The power of 31.5 from the original residuals (Figure 4) is greater
  than that from all 500 trials, implying an FAP for the 260 d period
  of less than 0.002, suggesting that it is real.}
\label{fig6}
\end{figure}
\clearpage

\begin{figure}
\plotone{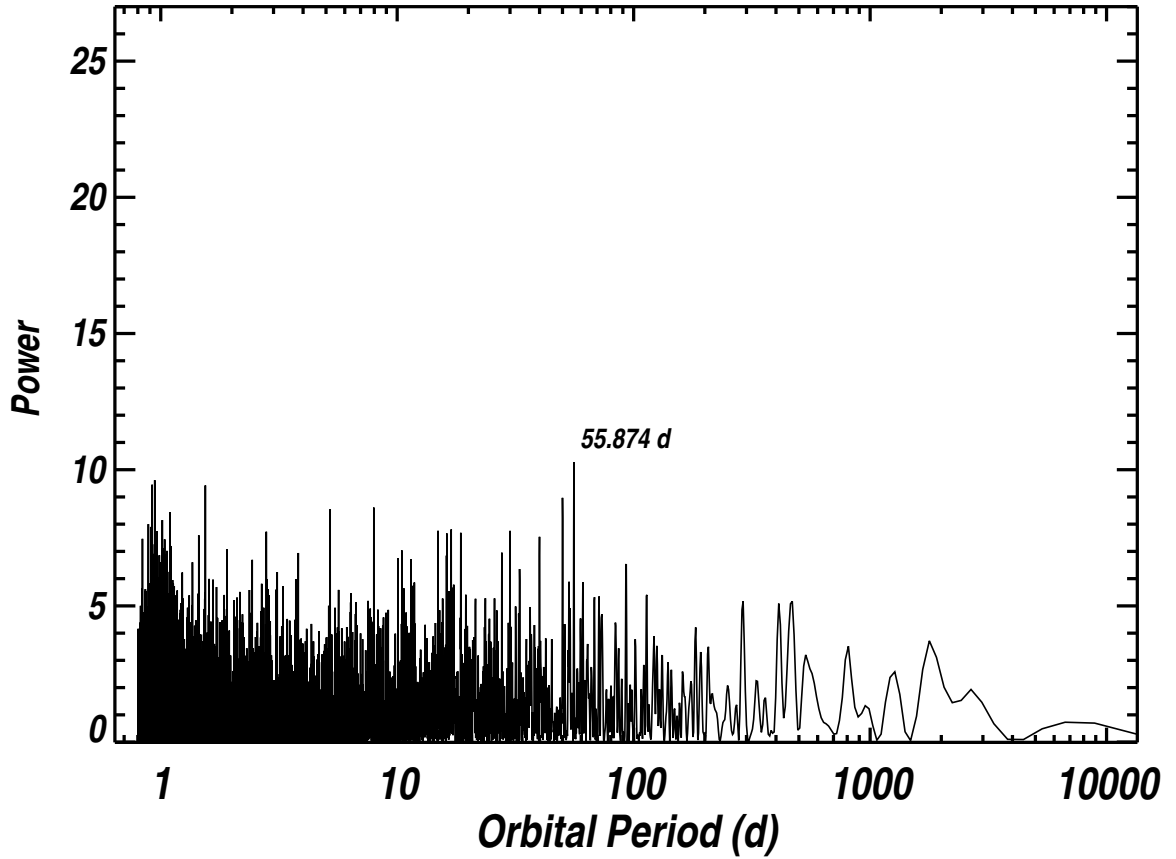}
\caption{Periodogram of the velocity residuals to a Keplerian model
  that contains five planets including the prospective new planet at $P
  \approx$ 260 d. 
  Velocities from both Lick and Keck are included.  
  The peak that had been apparent at 260 d in the residuals
  to a 4-planet model (Figures 4 and 5) has vanished due to
  its inclusion in the 5-planet model.  No other compelling
  periods are apparent.}
\label{fig7}
\end{figure}
\clearpage

\begin{figure}
\plotone{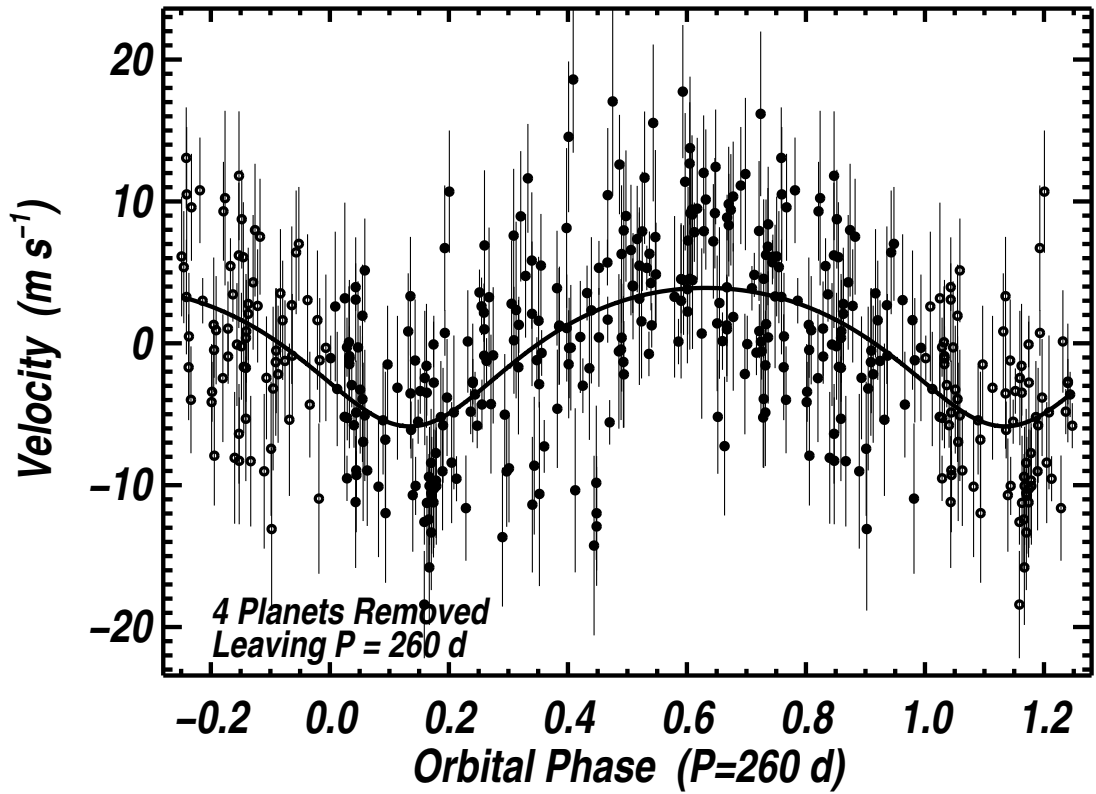}
\caption{Residual velocities vs orbital phase (for P = 260 d) 
 after the velocities induced by the four other planets are
  subtracted.  The orbital parameters were established with a
  simultaneous 5-planet Keplerian fit to all Doppler measurements.
  The residual velocities reveal the periodic variation
  associated with the new planet.  The solid line shows the Keplerian
  curve of the 260 d planet alone, with eccentricity frozen to 0.2.  
The planet's minimum mass is 
  45 \mearth and the semimajor axis is 0.78 AU.}
\label{fig8}
\end{figure}
\clearpage

\begin{figure}
\plotone{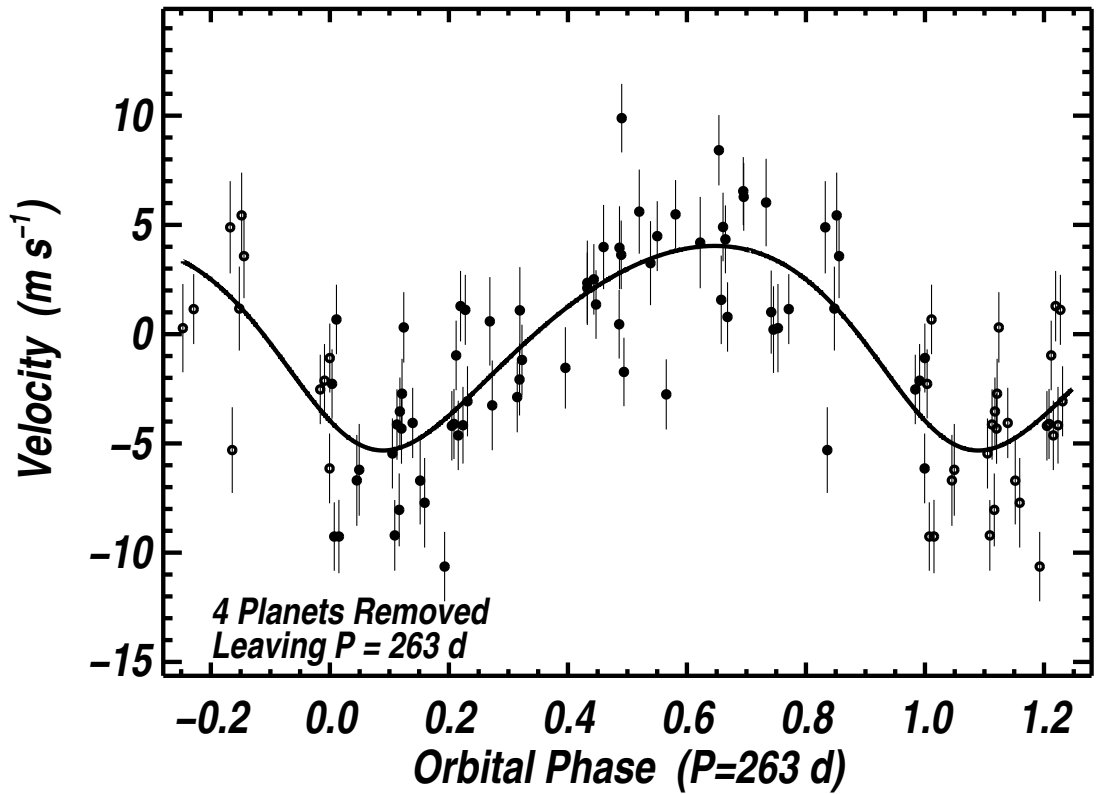}
\caption{Velocity of the new planet vs. orbital phase, as with Figure 8, 
  but for Keck velocities only.   The periodicity near 260 d is apparent
  independently in the Keck velocities.  The best-fit to the Keck
  velocities yields an eccentricity of $e$=0.16, only
  marginally significant.}
\label{fig9}
\end{figure}
\clearpage

\begin{figure}
\epsscale{1.0}
\plotone{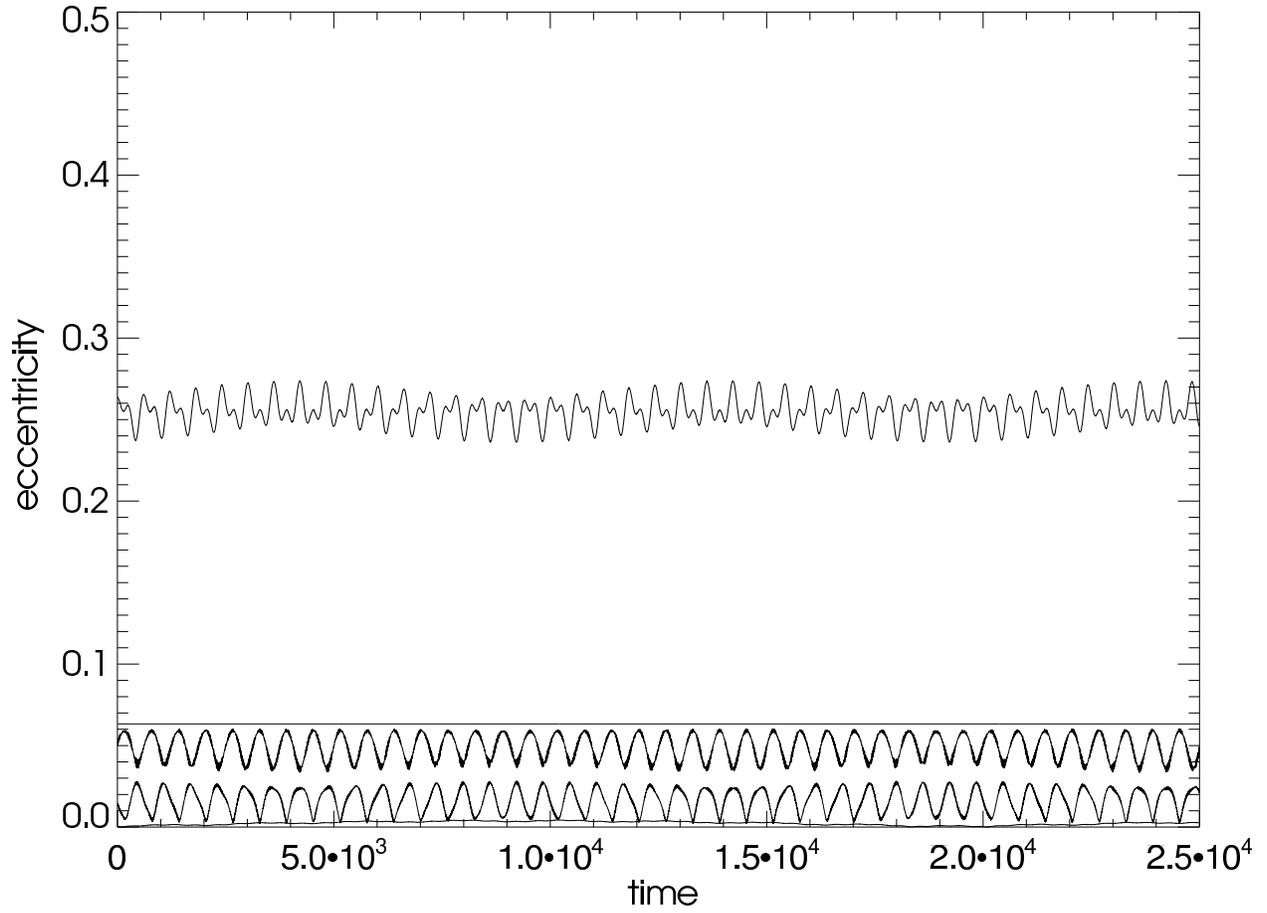}
\caption{
Eccentricity variations arising from an N-body numerical integration of the five-planet
model listed in Table 2.
\label{fig10}}
\end{figure}
\clearpage

\begin{figure}[t!]
\epsscale{0.85}
\plotone{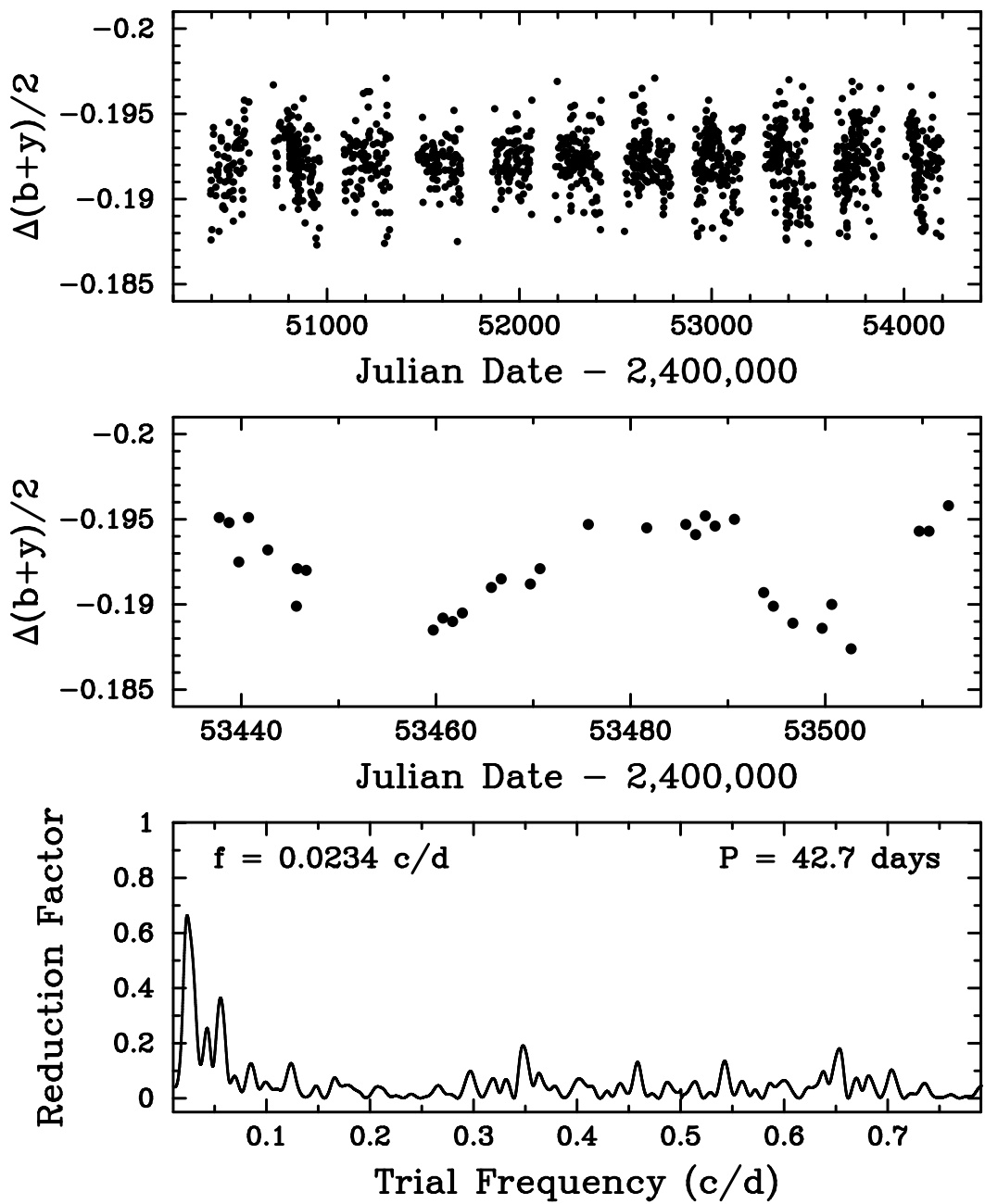}
\caption{The 11-yr Str\"omgren photometric data set of 55~Cnc ({\it top 
panel}).  The data have been normalized so that the annual means are
identical.  A portion of the ninth observing season ({\it middle panel})
shows coherent photometric variability in 55~Cnc due to rotational
modulation in the visibility of starspots.  The power spectrum of the
data in the middle panel ({\it bottom panel}) reveals the star's
rotation period of 42.7 $\pm$ 2.5 days.} 
\label{fig11}
\end{figure}

\begin{figure}[t!]
\epsscale{0.7} 
\plotone{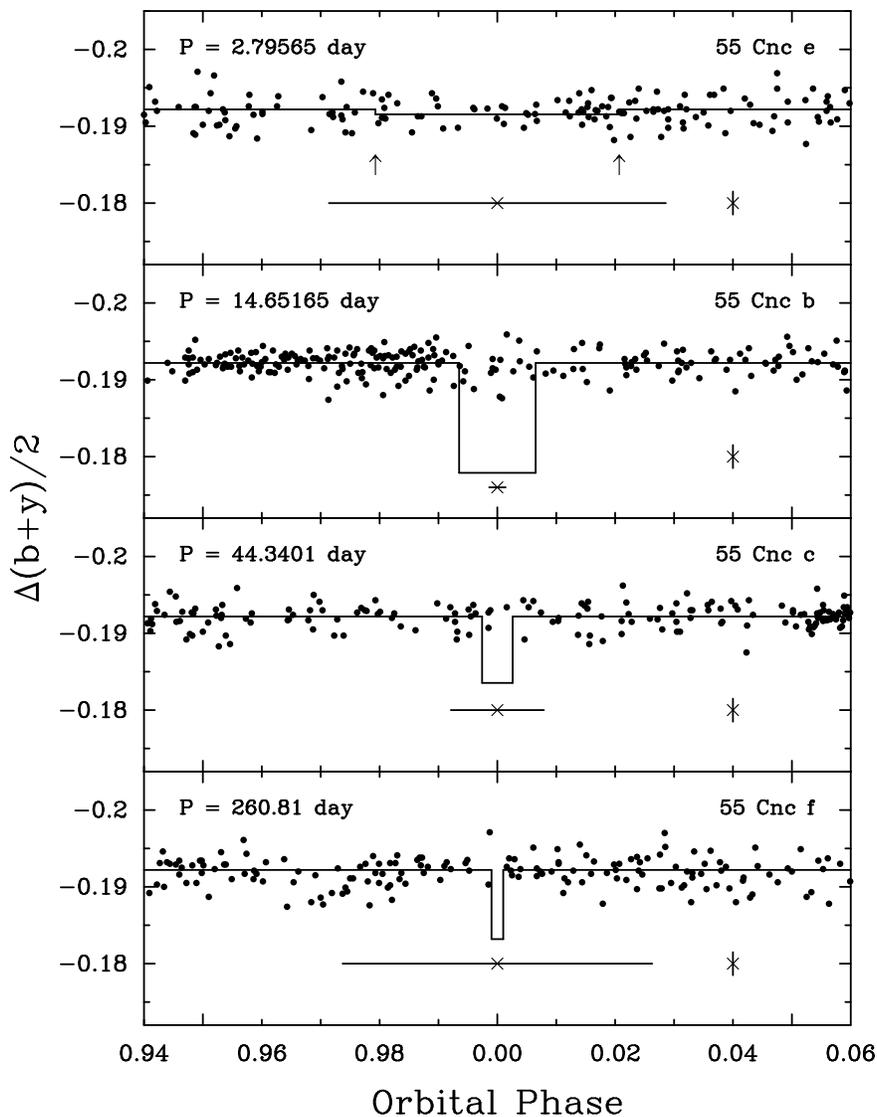}
\caption{Photometric observations of 55~Cnc plotted vs. orbital
phase for the inner four planetary companions, all plotted to the same
scale.  The solid line in each panel approximates the predicted
transit light curve, including the depth, duration, and timing of the
transits.  The arrows in the top panel indicate the beginning and end
of the very shallow predicted transits of the inner planet.  The
horizontal error bar beneath each transit box indicates the
uncertainty in the time of transit, while the vertical error bar shows
the nominal precision of a single data point.  The phase-folded
photometry does not detect transits for any of the four inner
planets. }
\label{fig12}
\end{figure}

\end{document}